\documentclass[12pt]{JHEP3}
\usepackage{mathrsfs}
\usepackage{amsmath,amssymb}
\usepackage{epsfig}
\usepackage{relsize}
\usepackage{bbm} 			% Fuer Matrix-1
\usepackage{dcolumn}  			% fuer ausrichtung innerhalb von Tabellen

\newcommand{\AdS}{\textrm{AdS}_5\times \textrm{S}^5}
\newcommand{\beq}{\begin{equation}}
\newcommand{\eeq}{\end{equation}}
\newcommand{\be}{\begin{equation}}
\newcommand{\ee}{\end{equation}}
\newcommand{\bea}{\begin{eqnarray}}
\newcommand{\eea}{\end{eqnarray}}
\newcommand{\ba}{\begin{array}}
\newcommand{\ea}{\end{array}}

\newcommand{\bra}[1]{\langle{#1}|}
\newcommand{\ket}[1]{|{#1}\rangle}

\newcommand{\nn}{\nonumber}

\newcommand{\hs}[1] {\hspace{#1}}
\newcommand{\vs}[1] {\vspace{#1}}

\newcommand{\tn}[1]{\textnormal{#1}}

\newcommand{\ns} \normalsize

\author{Alexander Hentschel, Jan Plefka and Per Sundin \\ \\ 
 {\it Humboldt-Universit\"at zu
Berlin, Institut f\"ur Physik,\\ Newtonstra{\ss}e 15, D-12489
Berlin, Germany}\\
{\tt alexander.hentschel,jan.plefka,per.sundin@physik.hu-berlin.de}\\} 
\abstract{We perform a detailed test of the quantum integrability of the $AdS_5\times S^5$ superstring in uniform 
light-cone gauge in its near plane-wave limit. 
For this we establish the form of the general nested light-cone Bethe equations 
for the quantum string from the long range $\mathfrak{psu(2,2|4)}$ Bethe equations of Beisert and Staudacher. 
Moreover the scheme for translating excited string states into Bethe root excitations is given. 
We then confront the direct perturbative diagonalization of
the light-cone string Hamiltonian in the near plane-wave limit with the energy spectrum obtained from the general 
nested light-cone Bethe equations in various higher rank sectors. The analysis is performed both analytically and numerically up to
the level of six impurity states and subsectors of maximal rank four. 
We find perfect agreement in all cases lending strong support to the quantum integrability of the $AdS_5\times S^5$ superstring.}

\title{Testing the nested light-cone Bethe equations of the \mathversion{bold}$AdS_5\times S^5$\mathversion{normal} 
superstring}
\preprint{HU-EP-07/07}
\begin{document}
\section{Introduction}

Determining the spectrum of the type IIB superstring on the maximally supersymmetric  
$AdS_5\times S^5$ background \cite{Metsaev:1998it} is of great interest, both in view of the AdS/CFT correspondence
\cite{ADSCFT} and as a problem in its own right within string theory. The string spectrum should be equivalent to the  
spectrum of scaling dimensions of local composite operators in the dual ${\cal N}=4$, $U(N)$ super Yang-Mills
theory in the 't Hooft limit.

In the last four years tremendous progress on this question has been made upon exploiting the 
{\sl assumed} property of integrablility in the system, following the pioneering work of Minahan and Zarembo 
\cite{MZ}\footnote{For reviews see \cite{revs}.}. 
Here progress was largely driven by advances on the gauge theory side, where it is possible
to map the perturbative spectral problem to the diagonalization of a corresponding super spin chain \cite{various}. 
Building upon one-loop studies \cite{Beisert:2005di} this finally led to the construction of a set 
of nested, asymptotic all-loop Bethe 
equations for the full model \cite{longrange}. Moreover the underlying symmetry
of the supergroup $PSU(2,2|4)$ was shown to determine the S-matrix \cite{Staudacher:2004tk} 
of the system up to an overall phase or 
dressing factor \cite{B0511}. 
As argued by Janik this abelian dressing factor can be constrained by crossing-invariance \cite{Janik} 
pointing towards an underlying Hopf algebraic structure \cite{Hopf}. 
Recently a proposal for the full dressing factor was made \cite{BHLBES} which remarkably agrees
with the findings of an independent four loop computation \cite{4loops} in the gauge theory.

Compared to these advances our understanding of the string side of the correspondence is less developed to date.
The sigma-model describing the $AdS_5\times S^5$ string is an integrable model \cite{Bena:2003wd}
at the classical level and one certainly hopes this to remain true also in the quantum theory. 
In \cite{Hofman:2006xt} a solitonic solution of the classical string was identified as the dual object to
the spin chain magnon, reproducing the spin chain dispersion relation in the strong t'Hooft coupling limit.
While it is unclear at present how to attack an exact quantization of the $AdS_5\times S^5$ string,
the problem is feasible upon consideration of suitable limits of the background geometry and perturbative 
expansions around them.
The most prominent example is the Penrose limit to a plane-wave background \cite{BMN}, where the string sigma
model becomes a free massive theory on the worldsheet. 
Here the first corrections to the plane-wave geometry can be treated
perturbatively and the leading corrections to the plane-wave spectrum was established in a series of papers
\cite{Parnachev,Callanetal,N_I}.
Moreover Arutyunov, Frolov and Staudacher \cite{AFS} showed that these corrections are reproduced from a set of 
quantum string Bethe equations in certain rank one subsectors, which have also been generalized to the full
model in \cite{longrange}. A central question in the analysis of the $AdS_5\times S^5$ superstring is that of a
convenient gauge choice for the worldsheet diffeomorphisms and kappa symmetry.
Building upon previous studies in reduced subsectors \cite{ulc1,ulc2} it was realized in \cite{ulcgauge}
that a uniform light-cone gauge employing the sum and difference of the global time coordinate and an 
angle on the $S^5$ as light-cone coordinates, along with a suitable kappa-symmetry gauge, 
simplifies the problem
considerably. In that paper the exact light-cone Hamiltonian
of the $AdS_5\times S^5$ string was established and the near plane-wave limit was performed, i.e.~the limit 
of large light-cone momentum $P_+$ with $\sqrt{\lambda}/P_+$ held fixed. The resulting
corrections at leading order $1/P_+$ in the light-cone energy were established and a set of ``light-cone'' 
Bethe equations was proposed, which reproduced these energy shifts in the rank one subsectors 
$\mathfrak{su}(2)$, $\mathfrak{sl}(2)$ and $\mathfrak{su}(1|1)$. Curiously, the form of these Bethe equations 
is simpler than the gauge theory inspired ones \cite{AFS} in that they come with a dressing factor equal to unity. This
statement is expected to hold, of course, only modulo unexplored terms at higher order in $1/P_+$. 
The residual symmetry structure of the light-cone gauged superstring was investigated in \cite{offshell}
and in \cite{ZF} assuming integrability a Zamolodchikov-Fadeev algebra was introduced for the
superstring.

One aim of the present paper is to clarify the connection of the light-cone Bethe equations to the
``standard'' gauge theory inspired Bethe equations of \cite{AFS} and its generalization to the full
higher rank system \cite{longrange} including the latest dressing factor. The set of nested light-cone
Bethe equations for general excitations of the near plane-wave superstring is derived and the translation
scheme from string oscillator excitations to Bethe root excitations is given. The energy shifts obtained
from solving the nested light-cone Bethe equations is confronted with the results of an explicit diagonalization
of the interacting near-plane wave Hamiltonian at leading order perturbation theory. 
This analysis is performed in higher rank subsectors of $\mathfrak{su}(1|2)$, $\mathfrak{su}(1,1|2)$ and
$\mathfrak{su}(2|3)$ analytically for lower excitation numbers and numerically for up to six excitations.
Perfect agreement is found in all cases, thus constituting a strong check of the quantum integrability
of the $AdS_5\times S^5$ superstring. If true the spectrum of the $AdS_5\times S^5$ superstring 
-- at least in the long string limit
$P_+\gg 1$ with all orders in a $1/P_+$ expansion included -- should be given by the solutions of the
general nested light-cone Bethe equations augmented by the conjectured dressing phase of \cite{BHLBES}.

Our analysis is complementary to the direct computation of the worldsheet S-matrix reported in \cite{McLKRZ}, 
see also \cite{KloseZarembo}.
In \cite{McLKRZ} the emergence of the two particle S-matrix of Beisert 
\cite{B0511} at leading order in $1/P_+$ was confirmed,
which is known to lead to the nested asymptotic Bethe equations of \cite{longrange}. 
This finding is a necessary but not sufficient condition for the integrability of the quantum $AdS_5\times S^5$ 
superstring, which would imply factorization of multi-particle scattering and the absence of particle production. 
Indeed the factorization of three particle scattering in the bosonic sector was demonstrated in the S-matrix approach 
of \cite{McLKRZ}. Our paper now provides a stringent test of the factorization property in larger sectors 
and at higher particle excitation numbers.

The plan of the paper is as follows. We begin by recalling the necessary facts of the uniform light-cone gauged
$AdS_5\times S^5$ superstring in the near plane wave limit in chapter two. Chapter three is then devoted to the
derivation of the nested light-cone Bethe equations for the full excitation structure. 
Moreover we present a string oscillator/Dynkin node excitation dictionary to translate the string into
the Bethe equation language.
In chapter four we discuss the large $P_+$ limit of this set of nested equations and present the emerging 
coupled polynomial equations for the Bethe roots which need to be solved. 
Explicit solutions are carried out for a number of subsectors and impurity numbers up to six 
(both with distinct and confluent mode numbers) in chapter five. The computations on the string side 
have been relegated to the appendix. 

\section{The Superstring on AdS$_5\times$ S$^5$}
\subsection{Hamiltonian in uniform light-cone gauge}

In \cite{ulcgauge} an exact gauge fixed Lagrangian of the Green-Schwarz Superstring on an
$\tn{AdS}_5 \times \tn{S}^5$ background was constructed in the 
uniform light-cone gauge \cite{ulc1, ulc2}. 
The associated light-cone
Hamiltonian is given by $\mathcal{H} = -P_-$ where $P_{\pm}
:= J \pm E$. Here $J$ denotes the angular momentum on $\tn{S}^5$ and
$E$ the global space-time energy.

Due to its nonlinearity an exact quantization of this system is unknown, nevertheless
the Hamiltonian of \cite{ulcgauge} allows for a perturbative quantization in the near plane wave
limit, where $P_+$ is taken to be large with $\widetilde{\lambda} :=
\frac{4\lambda}{P_+^2}$ held fixed. Using this approach the quantized
perturbative Hamiltonian has been computed up to next-to-leading order in a $1/P_+$ expansion
\begin{align} \label{eq:Hamiltonian_expasion}
   \mathcal{H} =& \mathcal{H}_2 + \frac{1}{P_+}\mathcal{H}_4 + \mathcal{O}(P_+^{-2})
\end{align}
The dynamical fields are given by the transverse eight fermionic and eight bosonic
fields. We will use the following decomposition of the eight complex
bosonic fields $Z_a, Y_a$ and their corresponding canonical momenta
$P^z_a, P^y_a$ following the conventions in \cite{ulcgauge}
\begin{align} \label{eq:bosonic_field_decomposition} \notag
  Z_a(\tau,\sigma) = & \sum_n e^{in\sigma}Z_{a,n}(\tau) & P^z_a(\tau,\sigma) = &\sum_n e^{in\sigma}P^z_{a,n}(\tau) \\ \notag
  Z_{a,n} = & \frac{1}{i\sqrt{\omega_n}}(\beta^+_{a,n}-\beta^-_{5-a,-n}) & P^z_{a,n} = & \frac{\sqrt{\omega_n}}{2}(\beta^+_{a,n}+\beta^-_{5-a,-n}) 
 \\ \notag
  Y_a(\tau,\sigma) = & \sum_n e^{in\sigma}Y_{a,n}(\tau) & P^y_a(\tau,\sigma) = &\sum_n e^{in\sigma}P^y_{a,n}(\tau) \\ 
  Y_{a,n} = & \frac{1}{i\sqrt{\omega_n}}(\alpha^+_{a,n}-\alpha^-_{5-a,-n}) & P^y_{a,n}=& \frac{\sqrt{\omega_n}}{2}(\alpha^+_{a,n}+\alpha^-_{5-a,-n}) 
  \hs{2pt},
\end{align}
where the frequency $\omega_n$ is defined as
\begin{align}
  \omega_n = \sqrt{1+\widetilde{\lambda}\, n^2} \, .
\end{align}
The decomposition has been chosen so that the creation and annihilation operators obey canonical commutation relations
\begin{align} \label{eq:bosonic_commutators}
   [\alpha^-_{a,n},\alpha^+_{b,m} ] = \delta_{a,b} \hs{2pt}\delta_{n,m} = [\beta^-_{a,n},\beta^+_{b,m} ],
\end{align}
where $a \in \{1,2,3,4\}$ is the flavor index and $n,m$ are the mode numbers which are subject to the level matching condition
\begin{align} \label{levelmatching}
  \sum_{j=1}^{K_4} m_j =0 \, ,
\end{align}
where $K_4$ denotes the total number of excitations. The mode
decompositions for the fermions\footnote{For the sake of completeness
  the mode decomposition of the $\eta$-field is given in this
  chapter. It is not to be confused with the grading $\eta_1$,
  $\eta_2$, which are used in section
  \ref{par:Bethe_equations_in_uniform_light-cone_gauge} to describe
  different choices of Dynkin diagrams for $\mathfrak{psu}(2,2|4)$}
are:
\begin{align} \label{eq:fermionic_field_decomposition} \notag
 \eta(\tau,\sigma) = & \sum_n e^{in\sigma}\eta_{n}(\tau) &  \theta(\tau,\sigma) = & \sum_n e^{in\sigma}\theta_{n}(\tau) \\ 
 \eta_n = & f_n \eta^-_{-n} + ig_n\eta^+_{n} 		 &  \theta_n = & f_n \theta^-_{-n} + ig_n \theta^+_{n} \\
\tn{with} \hs{30pt} \eta^-_{k} =  \eta^-_{a,k}&\Gamma_{5-a}\hs{3pt},  \hs{10pt} \eta^+_{k} = \eta^+_{a,k}\Gamma_{a}\hs{3pt}, &
	  \theta^-_{k} = \eta^-_{a,k}&\Gamma_{5-a}\hs{3pt}, \hs{10pt}  \theta^+_{k} =\eta^+_{a,k}\Gamma_{a} \, .
\end{align}
Where the explicit representation of the Dirac matrices $\Gamma_a$ is given in \cite{ulcgauge}. The functions $f_m$ and $g_m$ above are defined as
\bea
\label{fg}
f_m=\sqrt{\frac{1}{2}(1+\frac{1}{\omega_m})}, \quad g_m=\frac{\kappa\, \sqrt{\tilde{\lambda}}m}{1+\omega_m}f_m.
\eea
Here $\kappa=\pm 1$ is the arbitrary relative sign between kinetic and Wess-Zumino term in the 
worldsheet action.
The anti-commutators between the fermionic mode operators are then
\begin{align} \label{eq:fermionic_commutators}
   \{\eta^-_{a,n},\eta^+_{b,m} \} = \delta_{a,b} \hs{2pt}\delta_{n,m} = \{ \theta^-_{a,n},\theta^+_{b,m} \}\hs{2pt}.
\end{align}
Using this oscillator representation, the leading order Hamiltonian becomes
\begin{align} \label{eq:Hamiltonian_zeroth_order}
  \mathcal{H}_{2} = \sum_n \omega_n (\theta^+_{a,n}\theta^-_{a,n}+\eta^+_{a,n}\eta^-_{a,n}+\beta^+_{a,n}\beta^-_{a,n}+\alpha^+_{a,n}\alpha^-_{a,n})
  \hs{2pt}.
\end{align}
The first order correction to this Hamiltonian is given by \cite{ulcgauge}
\begin{align} \label{hamiltonian}
   \mathcal{H}_4 &= \mathcal{H}_{bb} + \mathcal{H}_{bf} + \mathcal{H}_{ff}(\theta) - \mathcal{H}_{ff}(\eta) \\
   \label{eq:Hamiltonian_H_bf_plefka_notation} \tn{with} \hs{30pt}
 \mathcal{H}_{bb} =\frac{\widetilde{\lambda}}{4}  
    ( Y'_{5-a} & Y'_aZ_{5-b}Z_b-Y_{5-a}Y_aZ'_{5-b}Z'_b +Z'_{5-a}Z'_aZ_{5-b}Z_b-Y'_{5-a}Y'_aY_{5-b}Y_b) \\ \notag
\label{middlepart}
 \mathcal{H}_{bf} = \frac{\widetilde{\lambda}}{4} \tn{tr} \Big[ \hs{10pt}	
  &(Z_{5-a}Z_a - Y_{5-a}Y_a)(\eta'^\dagger\eta' + \theta'^\dagger\theta')  \\ \notag  
 -&Z'_aZ_b[\Gamma_a,\Gamma_b]\left(	\mathcal{P}_+(\eta\eta'^\dagger - \eta'\eta^\dagger)- 
 					\mathcal{P}_-(\theta^\dagger\theta'-\theta'^\dagger\theta)\right)\\ \notag
 +&Y'_aY'_b[\Gamma_a,\Gamma_b]\left(	-\mathcal{P}_-(\eta^\dagger\eta' - \eta'^\dagger\eta)- 
 					 \mathcal{P}_+(\theta\theta'^\dagger-\theta'\theta^\dagger)\right) \\ \notag
 -& \frac{i\kappa}{\sqrt{\widetilde{\lambda}}}(Z_a P^z_b)'[\Gamma_a,\Gamma_b]
 	\left(  \mathcal{P}_+(\eta^\dagger \eta^\dagger + \eta\eta)+ \mathcal{P}_-(\theta^\dagger\theta^\dagger+\theta\theta)\right)\\ \notag
 +& \frac{i\kappa}{\sqrt{\widetilde{\lambda}}}(Y_a P^y_b)'[\Gamma_a,\Gamma_b]
 	\left(  \mathcal{P}_-(\eta^\dagger \eta^\dagger + \eta\eta)+ \mathcal{P}_+(\theta^\dagger\theta^\dagger+\theta\theta)\right)\\
 +& 8i Z_a Y_b \left(-\mathcal{P}_- \Gamma_a \eta' \Gamma_b \theta'  + \mathcal{P}_+ \Gamma_a\theta'^\dagger\Gamma_b\eta'^\dagger\right) 
 \hs{5pt} \Big]	   \\ \label{eq:Hamiltonian_H_ff_plefka_notation}
 \mathcal{H}_{ff}(\eta) = \frac{\widetilde{\lambda}}{4} \tn{tr} \Big[\Gamma_5 &\left(\eta'^\dagger\eta\eta'^\dagger\eta  
    + \eta^\dagger\eta'\eta^\dagger\eta' + \eta'^\dagger\eta^\dagger\eta'^\dagger\eta^\dagger + \eta'\eta\eta'\eta \right) \Big] \, .	  
\end{align}
This is the Hamiltonian for which we will determine the energy shifts $\delta P_-$
of the free, degenerate eigenstates $|\psi_{0,n}\rangle$ with 
${\cal H}_2\, |\psi_{0,n}\rangle = -(P_-)_0\, |\psi_{0,n}\rangle$
by diagonalizing the matrix $\langle\psi_{0,n}|{\cal H}_4|\psi_{0,m}\rangle$.
These will then be compared to the energies resulting from the proposed light-cone Bethe equations.
Due to the complexity of the Hamiltonian it is often
hard to obtain analytical results for these energy shifts in larger sectors with more than a few
number of excitations. We will then have to resort to numerical considerations.

%%%%%%%%%%%%%%%%%%%%%%%%%%%%%%%%%%%%%%%%%%%%%%%%%%%%%%%%%%%%%%%%%%%%%%%%%%%
\section{The light-cone Bethe equations for general sectors\label{par:Bethe_equations_in_uniform_light-cone_gauge}}
%%%%%%%%%%%%%%%%%%%%%%%%%%%%%%%%%%%%%%%%%%%%%%%%%%%%%%%%%%%%%%%%%%%%%%%%%%%

In an inspiring paper \cite{longrange} the long range gauge and string theory Bethe
equations were proposed for the full $\mathfrak{psu}(2,2|4)$ sector. This proposal was based
on a coordinate space, nested Bethe ansatz of the smaller $\mathfrak{su}(1,1|2)$ sector, 
a construction later on \cite{B0511} generalized to $\mathfrak{su}(2|3)$. See
\cite{Martins:2007hb} for a recent study of the problem employing the algebraic Bethe
ansatz. 
We shall start our analysis from the full set of  $\mathfrak{psu}(2,2|4)$ Bethe equations
proposed in \cite{longrange} in table 5 and adapt them to a language suitable for the 
light-cone gauge and large $P_+$ expansion performed in string theory \cite{ulcgauge}. 
This will set the basis for the subsequent comparison to the explicit diagonalization 
of the worldsheet Hamiltonian (\ref{hamiltonian}). 

The proposed set of Bethe equations for the spectral parameters $x_{i,k}$ of Beisert and Staudacher \cite{longrange} 
for the full model can be brought into the form
\begin{align}
\label{momentumcond}
1 &=\prod_{j=1}^{K_4} \frac{x^+_{4,k}}{x^-_{4,k}}\\
\label{const1}
1&=\prod_{j=1 \atop j\neq k}^{K_2} \frac{u_{2,k}-u_{2,j}-i\eta_1}{u_{2,k}-u_{2,j}+i\eta_1}
\prod_{j=1}^{K_3+K_1}\frac{u_{2,k}-u_{3,j}+\frac{i}{2}\eta_1}{u_{2,k}-u_{3,j}-\frac{i}{2}\eta_1} \\ 
  1&=\prod_{j=1}^{K_2}\frac{u_{3,k}-u_{2,j}+\frac{i}{2}\eta_1}{u_{3,k}-u_{2,j}-\frac{i}{2}\eta_1}\prod_{j=1}^{K_4}\frac{x_{4,j}^{+\eta_1}-x_{3,k}}{x_{4,j}^{-\eta_1}-x_{3,k}} \label{const1.5}
\\
\label{bethe}
1&=\Big(\frac{x^-_{4,k}}{x^+_{4,k}}\Big)^{L-\eta_1 K_1-\eta_2
  K_7}\, \prod_{j=1 \atop j\neq k}^{K_4}\Big(
\frac{x_{4,k}^{+\eta_1}-x_{4,j}^{-\eta_1}}{x_{4,k}^{-\eta_2}-x_{4,j}^{+\eta_2}}
\frac{1-g^2/(x^+_{4,k}x^-_{4,j})}{1-g^2/(x^-_{4,k}x^+_{4,j})}S^2_0
\Big)\nn\\ &\qquad\qquad\qquad\qquad \times
\prod_{j=1}^{K_3+K_1}\frac{x_{4,k}^{-\eta_1}-x_{3,j}}{x_{4,k}^{+\eta_1}-x_{3,j}}
\prod_{j=1}^{K_5+K_7}\frac{x_{4,k}^{-\eta_2}-x_{5,j}}{x_{4,k}^{+\eta_2}-x_{5,j}} \\
\label{const2}
1&=\prod_{j=1}^{K_6}\frac{u_{5,k}-u_{6,j}+\frac{i}{2}\eta_2}{u_{5,k}-u_{6,j}-\frac{i}{2}\eta_2}
\prod_{j=1}^{K_4}\frac{x_{4,j}^{+\eta_2}-x_{5,k}}{x_{4,j}^{-\eta_2}-x_{5,k}}\\
1&=\prod_{j=1 \atop j\neq k}^{K_6}
\frac{u_{6,k}-u_{6,j}-i\eta_2}{u_{6,k}-u_{6,j}+i\eta_2}\prod_{j=1}^{K_5+K_7}
\frac{u_{6,k}-u_{5,j}+\frac{i}{2}\eta_2}{u_{6,k}-u_{5,j}-\frac{i}{2}\eta_2}\, \label{last}.
\end{align}
In the above the variables $u_{i,k}$ are defined by $u_{i,k}=x_{i,k}+g^2\frac{1}{x_{i,k}}$
and the Bethe roots $x_{n,k}$ come with the multiplicities
\begin{align}
&x_{2,k}:\, k=1,\ldots, K_2 \qquad x_{3,k}:\, k=1,\ldots, (K_1+K_3)& \qquad x_{4,k}^\pm:\, k=1,\ldots K_4\nn\\
&x_{5,k}:\, k=1,\ldots, (K_5+K_7) \qquad x_{6,k}:\, k=1,\ldots, K_6& \qquad \phantom {x_{4,k_4}^\pm} 
\end{align}
Moreover the spectral parameters $x^\pm_{4,k}$ are related to the magnon momenta $p_k$ 
via
\be
\label{spectral} 
x^\pm_{4,k}=\frac{1}{4}(\cot\frac{p_k}{2}\pm i )
\Big(1+\sqrt{1+\frac{\lambda}{\pi^2}\sin^2\frac{p_k}{2}}\Big). 
\ee
and coupling constant $g^2$ is given by
\be
g:=\frac{\sqrt{\lambda}}{4\pi}=\frac{\sqrt{\tilde{\lambda}}P_+}{8 \pi} \, .
\label{gdef}
\ee
Note that we have chosen to write down the Bethe equations in a
more compact ``dynamically'' transformed language. In order to convert (\ref{momentumcond})-(\ref{last})
to the form found in table 5 of Beisert and Staudacher \cite{longrange} one introduces the $K_1$ resp. $K_7$ roots
$x_{1,k}$ and $x_{7,k}$ by splitting off the `upper' $x_{3,k}$ and $x_{5,k}$ roots via
\be
x_{1,k}:=g^2/ x_{3,K_3+k} \quad k=1,\ldots K_1 \qquad 
x_{7,k}:=g^2/ x_{5,K_5+k} \quad k=1,\ldots K_7 \, . 
\ee
This coordinate renaming unfolds the equations associated to the fermionic roots (\ref{const1}) and (\ref{const2}) into two 
structurally new sets of $K_1$ and $K_7$ equations and removes the $K_1$ and $K_7$ dependent exponent in
the central equation (\ref{bethe}).

The first equation (\ref{momentumcond}) of the form we will be using 
is the cyclicity constraint on the total momentum of the spin chain.
The following $K_2+(K_1+K_3)+K_4+(K_5+K_7)+K_6$ equations  in (\ref{const1})-(\ref{last}) 
determine the sets of Bethe roots $\{x_{2,k},x_{3,k},x^\pm_{4,k},x_{5,k},x_{6,k}\}$.
Let us stress once more that it is only the combinations $(K_1+K_3)$ and $(K_5+K_7)$ which enter in the
Bethe equations. Moreover the gradings 
 $\eta_1$ and $\eta_2$ take the values $\pm 1$
corresponding to four different choices of Dynkin diagrams for
$\mathfrak{psu}(2,2|4)$ as discussed in \cite{longrange} see figure 1. 

\begin{figure}\centering
$\{\eta_1,\eta_2\}=\{+1,+1\}$:\quad
\begin{minipage}{260pt}
\setlength{\unitlength}{1pt}%
\small\thicklines%
\begin{picture}(260,20)(-10,-10)
\put(  0,00){\circle{15}}%
\put(  -5,15){$K_1$}%
\put(  7,00){\line(1,0){26}}%
\put( 40,00){\circle{15}}%
\put(  35,15){$K_2$}%
\put( 47,00){\line(1,0){26}}%
\put( 80,00){\circle{15}}%
\put(  75,15){$K_3$}%
\put( 87,00){\line(1,0){26}}%
\put(120,00){\circle{15}}%
\put( 115,15){$K_4$}%
\put(127,00){\line(1,0){26}}%
\put(160,00){\circle{15}}%
\put(155,15){$K_5$}%
\put(167,00){\line(1,0){26}}%
\put(200,00){\circle{15}}%
\put(195,15){$K_6$}%
\put(207,00){\line(1,0){26}}%
\put(240,00){\circle{15}}%
\put(235,15){$K_7$}%
\put( -5,-5){\line(1, 1){10}}%
\put( -5, 5){\line(1,-1){10}}%
\put( 75,-5){\line(1, 1){10}}%
\put( 75, 5){\line(1,-1){10}}%
\put(155,-5){\line(1, 1){10}}%
\put(155, 5){\line(1,-1){10}}%
\put(235,-5){\line(1, 1){10}}%
\put(235, 5){\line(1,-1){10}}%
\put( 40,00){\makebox(0,0){$-$}}%
\put(120,00){\makebox(0,0){$+$}}%
\put(200,00){\makebox(0,0){$-$}}%
\end{picture}
\end{minipage}
\medskip\par
$\{\eta_1,\eta_2\}=\{+1,-1\}$:\quad
\begin{minipage}{260pt}
\setlength{\unitlength}{1pt}%
\small\thicklines%
\begin{picture}(260,20)(-10,-10)
\put(  0,00){\circle{15}}%
\put(  7,00){\line(1,0){26}}%
\put( 40,00){\circle{15}}%
\put( 47,00){\line(1,0){26}}%
\put( 80,00){\circle{15}}%
\put( 87,00){\line(1,0){26}}%
\put(120,00){\circle{15}}%
\put(127,00){\line(1,0){26}}%
\put(160,00){\circle{15}}%
\put(167,00){\line(1,0){26}}%
\put(200,00){\circle{15}}%
\put(207,00){\line(1,0){26}}%
\put(240,00){\circle{15}}%
\put( -5,-5){\line(1, 1){10}}%
\put( -5, 5){\line(1,-1){10}}%
\put( 75,-5){\line(1, 1){10}}%
\put( 75, 5){\line(1,-1){10}}%
\put(115,-5){\line(1, 1){10}}%
\put(115, 5){\line(1,-1){10}}%
\put(155,-5){\line(1, 1){10}}%
\put(155, 5){\line(1,-1){10}}%
\put(235,-5){\line(1, 1){10}}%
\put(235, 5){\line(1,-1){10}}%
\put( 40,00){\makebox(0,0){$-$}}%
\put(200,00){\makebox(0,0){$+$}}%
\end{picture}
\end{minipage}
\medskip\par
$\{\eta_1,\eta_2\}=\{-1,+1\}$:\quad
\begin{minipage}{260pt}
\setlength{\unitlength}{1pt}%
\small\thicklines%
\begin{picture}(260,20)(-10,-10)
\put(  0,00){\circle{15}}%
\put(  7,00){\line(1,0){26}}%
\put( 40,00){\circle{15}}%
\put( 47,00){\line(1,0){26}}%
\put( 80,00){\circle{15}}%
\put( 87,00){\line(1,0){26}}%
\put(120,00){\circle{15}}%
\put(127,00){\line(1,0){26}}%
\put(160,00){\circle{15}}%
\put(167,00){\line(1,0){26}}%
\put(200,00){\circle{15}}%
\put(207,00){\line(1,0){26}}%
\put(240,00){\circle{15}}%
\put( -5,-5){\line(1, 1){10}}%
\put( -5, 5){\line(1,-1){10}}%
\put( 75,-5){\line(1, 1){10}}%
\put( 75, 5){\line(1,-1){10}}%
\put(115,-5){\line(1, 1){10}}%
\put(115, 5){\line(1,-1){10}}%
\put(155,-5){\line(1, 1){10}}%
\put(155, 5){\line(1,-1){10}}%
\put(235,-5){\line(1, 1){10}}%
\put(235, 5){\line(1,-1){10}}%
\put( 40,00){\makebox(0,0){$+$}}%
\put(200,00){\makebox(0,0){$-$}}%
\end{picture}
\end{minipage}
\medskip\par
$\{\eta_1,\eta_2\}=\{-1,-1\}$:\quad
\begin{minipage}{260pt}
\setlength{\unitlength}{1pt}%
\small\thicklines%
\begin{picture}(260,20)(-10,-10)
\put(  0,00){\circle{15}}%
\put(  7,00){\line(1,0){26}}%
\put( 40,00){\circle{15}}%
\put( 47,00){\line(1,0){26}}%
\put( 80,00){\circle{15}}%
\put( 87,00){\line(1,0){26}}%
\put(120,00){\circle{15}}%
\put(127,00){\line(1,0){26}}%
\put(160,00){\circle{15}}%
\put(167,00){\line(1,0){26}}%
\put(200,00){\circle{15}}%
\put(207,00){\line(1,0){26}}%
\put(240,00){\circle{15}}%
\put( -5,-5){\line(1, 1){10}}%
\put( -5, 5){\line(1,-1){10}}%
\put( 75,-5){\line(1, 1){10}}%
\put( 75, 5){\line(1,-1){10}}%
\put(155,-5){\line(1, 1){10}}%
\put(155, 5){\line(1,-1){10}}%
\put(235,-5){\line(1, 1){10}}%
\put(235, 5){\line(1,-1){10}}%
\put( 40,00){\makebox(0,0){$+$}}%
\put(120,00){\makebox(0,0){$-$}}%
\put(200,00){\makebox(0,0){$+$}}%
\end{picture}
\end{minipage}
\caption{Four different choices of Dynkin diagrams of $\mathfrak{su}(2,2|4)$ 
specified by the grading $\eta_1$ and $\eta_2$. The signs in the white nodes indicate
the sign of the diagonal elements of the Cartan matrix \cite{longrange}.}
\end{figure}
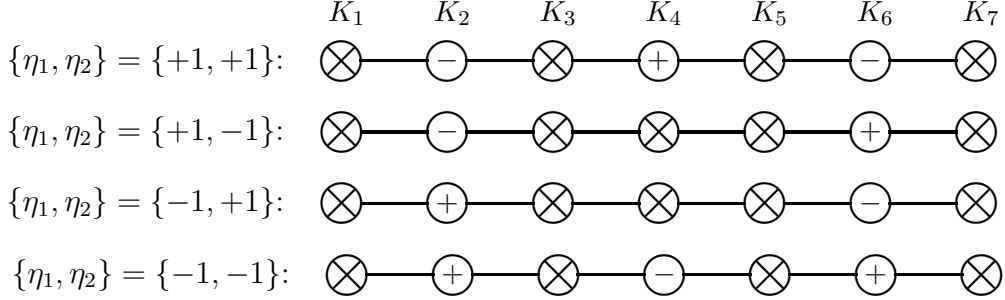

These four different choices of diagrams can
be traced back to the derivation of the nested Bethe ansatz in the
$\mathfrak{su}(1,1|2)$ sector in the gauge theory spin chain language. In this sector 
there are four distinct
excitations placed on a vacuum of $\mathcal{Z}$ fields. These four
excitations are the two bosonic $\mathcal{X}$ and $\mathcal{DZ}$ fields
and the two fermionic $\mathcal{U}$ and $\dot{\mathcal{U}}$ fields. In the
nested Bethe ansatz \cite{nested} one selects one out of these four
excitations as a second effective vacuum of a shorter spin chain,
after having eliminated all the sites $\mathcal{Z}$ from the original
chain. Depending on this choice $\eta_1$, $\eta_2$ take the values $\pm 1$.

Finally, the undetermined function $S_0^2$ in (\ref{bethe}) is the famous scalar dressing factor
which is conjectured to take the form
$S^2_0=S^2_0(x_{4,k},x_{4,j})=e^{2i\theta(x_{4,k},x_{4,j})}$ \cite{AFS}, where 
\be
\label{comb1}
\theta(x_{4,k},x_{4,j})= \sum_{r=2}^\infty \sum_{s=r+1}^\infty
c_{r,s}(g)\, \Bigl[ q_r(x^\pm_{4,k})\,q_s(x^\pm_{4,j}) -
q_r(x^\pm_{4,j})\,q_s(x^\pm_{4,k})\, \Bigr ] \ee with the local
conserved charge densities \be q_r(x^\pm) = \frac{i}{r-1}\, 
  g^{r-1}\, \left [\, \left (
    \frac{1}{x^+}\right )^{r-1} -\left ( \frac{1}{x^-}\right )^{r-1}\,
\right ] \ee and to leading order 
\begin{equation}
\label{leaddress}
c_{r,s}(g) = g\, \Bigl [\, \delta_{r+1,s} + {\cal O}(1/g)\, \Bigr ]\, .
\end{equation}
In this paper, we shall only be interested in
this leading order contribution, the AFS phase \cite{AFS}, where the phase factor may be summed \cite{Arutyunov:2006iu}
to yield 
\bea \theta_{kj} &= (x^+_j-x^+_k)\, F(x^+_kx^+_j)
+(x^-_j-x^-_k)\, F(x^-_k x^-_j) \nn\\\qquad & - (x^+_j-x^-_k)\,
F(x^-_k x^+_j) - (x^-_j-x^+_k)\, F(x^+_k x^-_j) \, , 
\eea 
with 
\be
F(a)=(1- \frac{g^2}{a})\, \log ( 1-\frac{g^2}{a})\, .  
\ee 
The string oscillator excitations are characterized by the values of four
$U(1)$ charges $(S_+,S_-,J_+,J_-)$ as introduced in \cite{offshell}.
They are related to the two spins $\{S_1,S_2\}$ on $AdS_5$ and two angular momenta 
$\{J_1,J_2\}$ on the $S_5$ via $S_\pm=S_1\pm S_2$ and $J_\pm=J_1\pm J_2$. The
relationship between these and
the excitation numbers $\{K_i\}$ in the Bethe equations are\footnote{To make a 
connection to \cite{longrange}, we
  have $J_-=q_1, J_+=q_2, S_-=s_1$ and $S_+=s_2$. The two other
  charges, $p$ and $r$ are functions of the length of the spin chain,
  so in the large
  $P_+$ limit these are infinite.} 
\bea
\label{dynkinlabels} \nn
&& S_+=\eta_2\, (K_5+K_7)-(1+\eta_2)\,K_6 +\frac{1}{2}(1-\eta_2)\, K_4, \\  \nn
&& S_-=\eta_1\,(K_1+K_3)-(1+\eta_1)\, K_2+\frac{1}{2}(1-\eta_1)\,K_4, \\ \nn
&& J_+=-\eta_2\, (K_5+K_7)-(1-\eta_2)\, K_6+\frac{1}{2}(1+\eta_2)\, K_4, \\ \nn
&& J_-=-\eta_1\, (K_1+K_3)-(1-\eta_1)\, K_2+\frac{1}{2}(1+\eta_1)\, K_4. 
\eea
Using these together with the $(S_+,S_-,J_+,J_-)$ charge values for the string oscillators of table
\ref{tab:fermionic_and_bosonic_field_charges} (see also \cite{offshell}) 
we can construct the excitation pattern for each oscillator, see table \ref{excpattern}.
\begin{table}[t]
\begin{center}
\begin{tabular}{|c||c|c|c|c|c||c|c|c|c|}
\hline
 & \textbf{K}$_1$ + \textbf{K}$_3$  & \textbf{K}$_2$   & \textbf{K}$_4$ & \textbf{K}$_6$ & \textbf{K}$_5$ 
+ \textbf{K}$_7$ & $S_+$ & $S_-$ & $J_+$ & $J_-$\\
\hline
$\alpha^+_1$ & $0+\frac{1}{2}(1-\eta_1)$ & 0 & 1 & 0 & $\frac{1}{2}(1-\eta_2)+0$   &0 &0 & 1 & 1\\

$\alpha^+_2$ & $\frac{1}{2}(1+\eta_1)+1$ & 1 & 1 & 0 & $\frac{1}{2}(1-\eta_2)+0$  &0 &0 & 1 & -1\\

$\alpha^+_3$ & $0+\frac{1}{2}(1-\eta_1)$ & 0 & 1 & 1 &$1+ \frac{1}{2}(1+\eta_2)$  &0 &0 & -1 & 1\\

$\alpha^+_4$ & $\frac{1}{2}(1+\eta_1)+1$ &1 &1 & 1 & $1+\frac{1}{2}(1+\eta_2)$  &0 &0& -1 & -1\\
\hline
$\beta^+_1$ & $0+\frac{1}{2}(1+\eta_1)$ & 0 & 1 &0  &$\frac{1}{2}(1+\eta_2)+0$   & 1 & 1& 0 & 0\\

$\beta^+_2$ & $\frac{1}{2}(1-\eta_1)+1$& 1 & 1 & 0 &$\frac{1}{2}(1+\eta_2)+0$  & 1 & -1 & 0 & 0 \\

$\beta^+_3$ & $0+\frac{1}{2}(1+\eta_1)$ & 0 &1 &1 &$1+ \frac{1}{2}(1-\eta_2)$   & -1 & 1& 0 & 0\\

$\beta^+_4$ & $\frac{1}{2}(1-\eta_1)+1$&1 &1 & 1 & $1+\frac{1}{2}(1-\eta_2)$   & -1 & -1& 0 & 0\\
\hline
$\theta^+_1$ & $0+\frac{1}{2}(1+\eta_1)$ & 0 & 1 & 0 & $\frac{1}{2}(1-\eta_2)+0$   & 0 & 1& 1 & 0\\

$\theta^+_2$ & $\frac{1}{2}(1-\eta_1)+1$ &1 &1 & 0 &$\frac{1}{2}(1-\eta_2)+0$   & 0 & -1& 1 & 0\\

$\theta^+_3$ & $0+\frac{1}{2}(1+\eta_1)$ & 0 &1 &1 & $1+\frac{1}{2}(1+\eta_2)$  & 0 & 1& -1 & 0\\

$\theta^+_4$ & $\frac{1}{2}(1-\eta_1)+1$&1 &1 & 1 & $1+\frac{1}{2}(1+\eta_2)$  & 0 & -1& -1 & 0\\
\hline
$\eta^+_1$ & $0+\frac{1}{2}(1-\eta_1)$ & 0 & 1 & 0 & $\frac{1}{2}(1+\eta_2)+0$  & 1 & 0& 0 & 1 \\

$\eta^+_2$ &$\frac{1}{2}(1+\eta_1)+1$ &1 &1 & 0 &$\frac{1}{2}(1+\eta_2)+0$   & 1 & 0 & 0 & -1 \\

$\eta^+_3$ & $0+\frac{1}{2}(1-\eta_1)$& 0 &1 &1 &$1+\frac{1}{2}(1-\eta_2)$   & -1 & 0& 0 & 1 \\

$\eta^+_4$ &$\frac{1}{2}(1+\eta_1)+1$ &1 &1 & 1 & $1+\frac{1}{2}(1-\eta_2)$   & -1 & 0& 0 & -1 \\
\hline
\end{tabular}
\caption{The translation scheme of string oscillator excitations to
  the Dynkin node excitation numbers of the Bethe equations. We have also listed
  the space-time $U(1)$ charges  $J_\pm$ and $S_\pm$ of the string oscillators. From this table
  we easily see which operators represent the middle node for the
  different choices of gradings. That is, $(\eta_1,\eta_1)=(+,+):
  \alpha^+_1,$ $(-,+):\theta^+_1,$ $(+,-):\eta^+_1$ and
  $(-,-):\beta^+_1$.}
\label{excpattern}
\end{center}
\end{table}
For example, the excitations in the $\mathfrak{su}(1,1|2)$ sector correspond to the following string oscillators,
\bea
\mathcal{X} \doteq \alpha^+_1, \quad \mathcal{DZ} \doteq \beta^+_1,
\quad \mathcal{U} \doteq \theta^+_1, \quad \mathcal{\dot{U}} \doteq
\eta^+_1.  \eea These are the four fields which are picked out as
a new vacuum in the smaller spin chains by specifying the
values\footnote{The field that is picked as the second vacuum in the nested Bethe ansatz only
  excites the middle node of the Dynkin diagram, so one immediately
  sees from the table which combinations of the gradings correspond to which choice of vacuum.} of the gradings $\eta_1$
and $\eta_2$. The vacuum of $\mathcal{Z}$ fields corresponds to the
string ground state $\ket{0}$ with charge $J$.

Let us stress that in the dictionary of table \ref{excpattern} a single string oscillator excitation 
does not corresponds to a single Dynkin node excitation, but rather to a five component excitation vector,
with uniform $K_4=1$ entry. This is how the naive mismatch of 16 string oscillators versus 7 (or better 4) 
Dynkin node excitations is resolved: One should think of a string oscillator as being indexed by the
space-time charge vector $(S_+,S_-,J_+,J_-)$ {\sl or} by the Dynkin vector $(K_1+K_3, K_2, K_6, K_5+K_7)$.
These two labelings are equivalent and the one-to-one map between them is given in (\ref{dynkinlabels}).

There are several things we need to do in order to translate the Bethe equations
(\ref{momentumcond})-(\ref{last}) into their light-cone form in order to make a direct
comparison to uniform light-cone gauged, near plane-wave string theory. First of all, since
the light-cone Hamiltonian is expanded in the large $P_+$ limit we
need to express $L$ in (\ref{bethe}) in terms of the light-cone
momenta. This can be done by using the expression for the eigenvalues
of the dilatation operator and the $J$ charge of $S^5$ \cite{longrange}, \bea
\label{DJ}
&& J=L+\frac{1}{2}\eta_1(K_3-K_1)-\frac{1}{4}(2+\eta_1+\eta_2)K_4+\frac{1}{2}\eta_2(K_5-K_7), \\ \nn
&& D=L+\frac{1}{2}\eta_1(K_3-K_1)+\frac{1}{4}(2-\eta_1-\eta_2)K_4+\frac{1}{2}\eta_2(K_5-K_7)+\delta D,
\eea
where the anomalous dimension $\delta D$ reads
\bea
\label{anamolous}
\delta D =2g^2\sum_{j=1}^{K_4}\Big(\frac{i}{x^+_{4,j}}-\frac{i}{x^-_{4,j}}\Big)\, ,
\eea 
Using (\ref{DJ}) we can write the light-cone momenta and energy as,
\begin{align}
  \label{lc}
  P_+ & = D+J \\ \nonumber
       & = 2L+\eta_1(K_3-K_1)-\frac{1}{2}(\eta_1+\eta_2)K_4+\eta_2(K_5-K_7)+\delta D \\ \nonumber
  P_- & = J-D=-K_4-\delta D \hspace{3pt}.
\end{align}
Hence we see that the large $P_+$ limit discussed in the previous section corresponds to an 
infinitely long chain with a finite number of excitations.  
Using this, the central $K_4$ Bethe equations (\ref{bethe}) become
\bea &&
\Big(\frac{x^+_{4,k}}{x^-_{4,k}}\Big)^{\frac{1}{2}P_+}=\Big(\frac{x^-_{4,k}}{x^+_{4,k}}\Big)^{\frac{1}{2}(\frac{1}{2}(\eta_1+\eta_2)K_4-\eta_1(K_1+K_3)-\eta_2(K_5+K_7)-\delta
  D)} \\ \nn && \times \prod_{j=1 \atop j\neq k}^{K_4} \Big(\frac{x_{4,k}^{+\eta_1}-x_{4,j}^{-\eta_1}}{x_{4,k}^{-\eta_2}-x_{4,j}^{+\eta_2}}\frac{1-g^2/(x^+_{4,k}x^-_{4,j})}{1-g^2/(x^-_{4,k}x^+_{4,j})}S^2_0\Big)
\prod_{j=1}^{K_3+K_1}\frac{x_{4,k}^{-\eta_1}-x_{3,j}}{x_{4,k}^{+\eta_1}-x_{3,j}}\prod_{j=1}^{K_5+K_7}\frac{x_{4,k}^{-\eta_2}-x_{5,j}}{x_{4,k}^{+\eta_2}-x_{5,j}}.
\eea We want to compare the spectrum up to
$\mathcal{O}(\frac{1}{P^2_+})$ and to this order a nice thing
happens. As a matter of fact, one can show using only the leading AFS piece of (\ref{leaddress}) that
\be
\label{combination}
\Big(\frac{x^-_{4,k}}{x^+_{4,k}}\Big)^{-\frac{1}{2}\delta D} \,
 \prod_{j=1 \atop j\neq k}^{K_4}\Big(\frac{1-g^2/(x^+_{4,k}x^-_{4,j})}{1-g^2/(x^-_{4,k}x^+_{4,j})}S^2_0\Big)
=1+{\cal O}(\frac{1}{P^3_+})
\ee 
holds, once one inserts the large $P_+$ expansion of $p_k$ (to be established in (\ref{momentum}) and (\ref{x4exp})) as well as the relevant leading AFS contribution to the dressing factor $S_0$ of (\ref{leaddress}).
Curiously enough, not only the $1/P_+$ contribution, but also the $1/P_+^2$ term vanishes in this expansion -- the 
$1/P_+^3$ term is nonvanishing though. 
Therefore, to the order we are interested in, the light-cone Bethe
equations are given by the previous equations of (\ref{momentumcond})-(\ref{last})
with the central node $K_4$ Bethe equations (\ref{bethe}) exchanged by the simpler dressing factor free form
\bea
\label{finalbethe}
&&
\Big(\frac{x^+_{4,k}}{x^-_{4,k}}\Big)^{\frac{1}{2}P_+}=
\Big(\frac{x^-_{4,k}}{x^+_{4,k}}\Big)^{\frac{1}{2}(\frac{1}{2}(\eta_1+\eta_2)K_4-\eta_1(K_1+K_3)-\eta_2(K_5+K_7))}
\\ \nn && \times \prod_{j=1 \atop j\neq k}^{K_4}
\frac{x_{4,k}^{+\eta_1}-x_{4,j}^{-\eta_1}}{x_{4,k}^{-\eta_2}-x_{4,j}^{+\eta_2}}
\prod_{j=1}^{K_3+K_1}\frac{x_{4,k}^{-\eta_1}-x_{3,j}}{x_{4,k}^{+\eta_1}-x_{3,j}}
\prod_{j=1}^{K_5+K_7}\frac{x_{4,k}^{-\eta_2}-x_{5,j}}{x_{4,k}^{+\eta_2}-x_{5,j}} +\mathcal{O}(\frac{1}{P_+^2}) \, ,
\eea 
Putting all $K_j=0$, for $j\neq 4$, we indeed reproduce the results for the rank one subsectors presented
in \cite{ulcgauge}. This explains the simple form of the equations established there.
%%%%%%%%%%%%%%%%%%%%%%%%%%%%%%%%%%%%%%%%%%%%%%%%%%%%%%%%%%%%%%%%%%%%%%%%%%%%%%%%%%%%%%%%%%%%%%%%%%%%%%%%%%%%%%%%%%%%
  \section{Large $P_+$ expansion}
%%%%%%%%%%%%%%%%%%%%%%%%%%%%%%%%%%%%%%%%%%%%%%%%%%%%%%%%%%%%%%%%%%%%%%%%%%%%%%%%%%%%%%%%%%%%%%%%%%%%%%%%%%%%%%%%%%%%

We will now explicitly expand the Bethe equations in the large $P_+$ limit. The mode numbers of the string oscillators will enter in the equations as the zero mode of the magnon momenta $p_k$. However, depending on if we are looking at a state with confluent mode numbers or not, the procedure is somewhat different. We will begin with the simpler case where all mode numbers are distinct.
\subsection{Non-confluent mode numbers}
For distinct mode numbers one assumes an expansion of $p_k$ as \cite{AFS,ulcgauge} 
\bea
\label{momentum}
p_k=\frac{p^0_k}{P_+}+\frac{p^1_k}{P^2_+}.  
\eea 
Determining the analogous expansion of $x^\pm_{4,k}$
\begin{equation}
x^\pm_{4,k}=P_+\, x^0_{4,k}+x^{1,\pm}_{4,k}+\ldots \, ,
\end{equation}
where
\bea
\label{x4exp}
x^0_{4,k}=\frac{1+\omega_k}{2p^0_k},\quad  x^{1,\pm}_{4,k}= \frac{1}{4}(1+\omega_k)\Big( \pm i - \frac{2 p^1_k}{(p^0_k)^2 \hspace{2pt} \omega_k} \Big) \hspace{3pt},
\eea
and $\omega_k=\sqrt{1+\tilde{\lambda}\frac{(p^0_k)^2}{16\pi^2}}$.
Consistency then
implies that the spectral parameters
$x_{3,k}$ and $x_{5,k}$ have the expansion\footnote{The expansion of
  $x_{3,k}$ and $x_{5,k}$ remains the same in the case of
  confluent mode numbers, while the expansion of $x^\pm_{4,k}$
  differs.}  
\bea
\label{rootexpansion}
x_{3,k}=P_+\, x^0_{3,k}+x^1_{3,k}+\ldots, \qquad x_{5,k}=P_+\, x^0_{5,k}+x^1_{5,k}+\ldots \, .
\eea
Taking the logarithm of (\ref{finalbethe}) and expanding we find that the momentum at leading order 
$p^0_k$ in (\ref{momentum}) satisfy
\bea
p^0_k=4\pi m_k, \qquad m_k \in \mathbb{Z},
\eea
the integer here is what will correspond to the mode numbers of the string oscillators. 
Expanding (\ref{finalbethe}) to the next order we find that the $p^1_k$ should satisfy
\begin{align}
\label{expandedbethe}
 p^1_k= \hspace{10pt} & \frac{1}{2}(\eta_1+\eta_2)\sum_{j=1 \atop j \neq k}^{K_4}\frac{2+\omega_{k}+\omega_{j}}{x^0_{4,k}-x^0_{4,j}}
                    -  \eta_1\sum_{j=1}^{K_1+K_3}\frac{1+\omega_{k}}{x^0_{4,k}-x^0_{3,j}} \\ \nn
                    -& \eta_2\sum_{j=1}^{K_5+K_7}\frac{1+\omega_{k}}{x^0_{4,k}-x^0_{5,j}} 
                    -  (\frac{1}{2}(\eta_1+\eta_2)K_4-\eta_1(K_1+K_3)-\eta_2(K_5+K _7))p^0_k.
\end{align}
We also want to expand the light-cone energy (\ref{lc}), using (\ref{anamolous}) and (\ref{spectral}) we find
\bea
P_-= - \sum_{k=1}^{K_4}\omega_k+\delta P_-, 
\eea
where the energy shift, $\delta P_-$, is given by
\bea
\label{energyshift}
\delta P_-= -\frac{\tilde{\lambda}}{P_+}\frac{1}{16 \pi^2}\sum_{k=1}^{K_4}\frac{p^0_k p^1_k}{\omega_k}.
\eea
\subsection{Confluent mode numbers \label{section:Confluent_mode_numbers}}
For the case of confluent mode numbers we run into trouble because of the zero denominator in (\ref{expandedbethe}), which is caused by the term
\bea
\label{problematic}
\prod_{j=1 \atop j \neq
  k}^{K_4}\frac{x_{4,k}^{+\eta_1}-x_{4,j}^{-\eta_1}}{x_{4,k}^{-\eta_2}-x_{4,j}^{+\eta_2}}
\eea 
of \eqref{finalbethe}. One could try to only look at the case with the gradings
chosen so that $\pm \eta_1=\mp \eta_2$. However, this would mean that
we pick a fermionic vacuum in the nested Bethe ansatz and since the
rapidities $x_{4,k}$ are degenerate, we end up with zero. So for the case of confluent mode numbers we are forced to pick $\eta_1=\eta_2$.

The way to proceed is to assume an expansion of $p_k$ as \cite{AFS},
\bea
\label{confluentmomenta}
p_k=\frac{p^0_k}{P_+}+\frac{p^1_{k,l_k}}{P_+^{3/2}}+\frac{p^2_{k,l_k}}{P_+^2}.
\eea Where we, following \cite{AFS}, denote the multiplicity of the
degeneracy as $\nu_k$ so $\sum_{k=1}^{K_4'}\nu_k=K_4$ and
$\sum_{k=1}^{K_4'}\nu_k m_k=0$, where $K_4'$ is the number of distinct
mode numbers. The first order term in (\ref{confluentmomenta}) is
degenerate for confluent mode numbers while for the higher order terms
the degeneracy might be lifted ($l_k \in \{1,2,...,\nu_k\}$).

Using (\ref{confluentmomenta}) the energy shift will decompose as
\bea
\delta P_-=\sum_{k=1}^{K_4'}\sum_{l_k=1}^{\nu_k}\delta P_{-_,k,l_k}.
\eea
The contribution from mode numbers $m_j$ with $\nu_j=1$ look the same as in (\ref{energyshift}) while modes $m_k$ with $\nu_k > 1$ will have contribution from $p^1_{k,l_k}$. Using (\ref{confluentmomenta}) and expanding (\ref{problematic}) we find that $p^1_{k,l_k}$ satisfy a Stieltjes equation \cite{stieltjes} of the form \cite{AFS}
\bea
\label{p1k}
p^1_{k,l_k}=-2(\eta_1+\eta_2) (p^0_k)^2\omega_{k}\sum_{\mu_k=1 \atop \mu_k\neq l_k}^{\nu_k}\frac{1}{p^1_{k,l_k}-p^1_{k,\mu_k}}.
\eea
It is useful to note that $\sum_{l_k=1}^{\nu_k}p^1_{k,l_k}=0$. The momenta $p^1_{k,l_k}$ can be written as
\bea
(p^1_{k,l_k})^2=-2\, (\eta_1+\eta_2)\, (p^0_k)^2\, \omega_k \, h_{\nu_k,l_k}^2 \hspace{20pt} \textnormal{with } \hspace{5pt} l_k = 1,...,\nu_k
\eea
where $h_{\nu_k,l_k}$ are the $\nu_k$ roots of Hermite polynomials of degree $\nu_k$. 
However, the explicit solutions $h_{\nu_k,l_k}$ are not needed since when summing over $k$ the following property applies 
\bea
\sum_{l_k=1}^{\nu_k}(h_{\nu_k,l_k})^2=\frac{\nu_k(\nu_k-1)}{2}.
\eea
The expansion for the second order contribution $p^2_{k,l_k}$ in (\ref{confluentmomenta}) is considerably more complicated, we therefore refer only to its general structure
\begin{align} 
  p^2_{k,l_k} = \widetilde{p}^{\hspace{2pt} 2}_{k} + \sum^{\nu_k}_{\mu_k =1 \atop \mu_k \neq l_k} f_k(\mu_k,l_k) \hspace{3pt}.
\end{align}
We split $p^2_{k,l_k}$ into a part not depending on $l_k$, which is equivalent to $p^1_k$ given in \eqref{expandedbethe}: $\widetilde{p}^{\hspace{2pt} 2}_{k} \equiv p_k^1$. The function $f_k$ has the property $f_k(\mu_k,l_k) = - f_k(l_k,\mu_k)$ and thus the second term drops out when summed over $l_k$. 
% sum over $k$ in the energy shift and effectively the only contributing part is identical to (\ref{expandedbethe})
% \footnote{So in this notation, $p^2_{k,l_k}$ correspond to $p^1_k$ in (\ref{expandedbethe}).}, 
% so for $\delta P_-$ we can remove the $m_k$ index from $p^2_k$. 
The final expression for the energy shift becomes then
\begin{align}
\label{expandedenergycoinciding}
 \delta P_-
   & = - \frac{1}{P_+}\frac{\tilde{\lambda}}{16\pi^2}\sum_{k=1}^{K_4'}
         \sum_{l_k=1}^{\nu_k}\frac{\frac{1}{2}(p^1_{k,l_k})^2+p^0_k \omega_k^2 p^2_{k,l_k}}{\omega_k^3} \\ \nn
   & = - \frac{1}{P_+}\frac{\tilde{\lambda}}{32\pi^2}\sum_{k=1}^{K_4'}\nu_k p^0_k 
         \Big(\frac{2 {\tilde p}^2_k\omega_k-(\eta_1+\eta_2)p^0_k(\nu_k-1)}{\omega_k^2}\Big).
\end{align}

%%%%%%%%%%%%%%%%%%%%%%%%%%%%%%%%%%%%%%%%%%%%%%%%%%%%%%%%%%%%%%%%%%%%%%%%%%%%%%%%%%%%%%%%%%%%%%%%%%%%%%%%%%%%%%%%%%%%%%%%%%%%%%%%%%%%%%%%%%%%%%%
   \subsection{Bethe equations for the smaller spin chains}
%%%%%%%%%%%%%%%%%%%%%%%%%%%%%%%%%%%%%%%%%%%%%%%%%%%%%%%%%%%%%%%%%%%%%%%%%%%%%%%%%%%%%%%%%%%%%%%%%%%%%%%%%%%%%%%%%%%%%%%%%%%%%%%%%%%%%%%%%%%%%%%
To be able to solve for $p_k^1$ it is clear from the form of (\ref{expandedbethe}) that we need the values of the Bethe
roots $x_{3,k}$ and $x_{5,k}$ at leading order in $P_+$. Note that the variables $u_k$ scale as
$u_k=P_+u^0_k+u^1_k+\ldots$. Expanding (\ref{const1}), (\ref{const1.5}), (\ref{const2}) and (\ref{last}) yields
\bea
&& 0=\sum_{j=1 \atop j\neq
  k}^{K_2}\frac{2}{u^0_{2,j}-u^0_{2,k}}+\sum_{j=1}^{K_1+K_3}\frac{1}{u^0_{2,k}-(x^0_{3,j}+\frac{\tilde{\lambda}}{64\pi^2}\frac{1}{x^0_{3,j}})}\, , \nn
\\ \nn &&
0=\eta_1\sum_{j=1}^{K_2}\frac{1}{x^0_{3,k}+\frac{\tilde{\lambda}}{64\pi^2}\frac{1}{x^0_{3,k}}-u^0_{2,j}}+\frac{1}{2}\sum_{j=1}^{K_4}\frac{1+\omega_j}{x^0_{4,j}-x^0_{3,k}}\, ,
\\ \nn &&
0=\eta_2\sum_{j=1}^{K_6}\frac{1}{x^0_{5,k}+\frac{\tilde{\lambda}}{64\pi^2}\frac{1}{x^0_{5,k}}-u^0_{6,j}}+\frac{1}{2}\sum_{j=1}^{K_4}\frac{1+\omega_j}{x^0_{4,j}-x^0_{5,k}}\, , \nn
\\ && 0=\sum_{j=1 \atop j\neq
  k}^{K_6}\frac{2}{u^0_{6,j}-u^0_{6,k}}+\sum_{j=1}^{K_5+K_7}\frac{1}{u^0_{6,k}-(x^0_{5,j}+\frac{\tilde{\lambda}}{64\pi^2}\frac{1}{x^0_{5,j}})}\, , \label{constexpanded}
\eea 
which determine the $x^0_{2,k}$, $x^0_{3,k}$, $x^0_{5,k}$ and $x^0_{6,k}$ in terms of $x^0_{4,k}$. Note that the two sets of
the first two and the last two equations are decoupled and identical in structure.

Let us briefly discuss how one goes about solving these equations for
a given excitation sector.
First one needs to commit oneself to a specific grading by specifying the numbers 
$\eta_{1,2}=\pm 1$. Then one reads off
the values for $\{K_i\}$ in table \ref{excpattern} corresponding to the
excitation pattern in question. The four different
choices of gradings can be grouped into two classes, one with
fermionic middle node, $\eta_1=-\eta_2$, and one with bosonic middle
node, $\eta_1=\eta_2$ in the associated Dynkin diagram. The difference between 
the two is important in the case of confluent
mode numbers. The $K_3$ and $K_5$ (and for $\eta_1=-\eta_2$,
also $K_4$) are fermionic nodes which means that the solutions for
$x^0_{3,k}$ and similarly for $x^0_{5,k}$ for different
values of $k$ are not allowed to be degenerate by the Pauli principle.
 
Consider for example the $\mathfrak{su}(1,1|2)$ sector containing
only nonvanishing values for $\{K_3,K_4,K_5\}$. Then, due to $K_2=0=K_6$, 
the equations (\ref{constexpanded}) condense to two identical, degree $K_4$
polynomial equations for $x_{3,k}^0$ and $x_{5,k}^0$ yielding
$K_4$ solutions, including the degenerate solution $\{ x^0_{3/5,k}\to \infty\}$.
These $K_4$ solutions are then used once on each node $K_3$ and
$K_5$, each generating
$\frac{K_4\, (K_4-1)\times...\times(K_4-K_{j})}{K_{j}!}$ (with $j=3,5$)
number of solutions. For a bosonic node, however, we may pick the same
solution repeatedly. 

Having distributed the solutions for $x^0_{3,k}$ and $x^0_{5,k}$ one then
determines $p^1_k$ from (\ref{expandedbethe}) and finally solves for the
energy shift using (\ref{energyshift}) or (\ref{expandedenergycoinciding}). The 
obtained value is what we then compare with a direct
diagonalization of the string Hamiltonian.

\section{Comparing the Bethe equations with string theory}

We have calculated the energy shifts (both analytically and numerically) for a
large number of states. The numerical results will be presented in
appendix \ref{par:Numerics}, while here in the main text we shall focus on the analytical
results. On the string theory side one studies the Hamiltonian in first order degenerate
perturbation theory, which in practice demands the diagonalization of the
Hamiltonian in the relevant subsectors. 
In the near plane-wave limit, this was first done 
in \cite{Paranachev,Callanetal} using a different gauge.

\subsection{General structure of solutions}

We will present analytical results for three different sectors,
$\mathfrak{su}(1|2)$, $\mathfrak{su}(1,1|2)$ and $\mathfrak{su}(2|3)$.
The operators in each sector are \bea \mathfrak{su}(1|2): \quad
\{\alpha^+_1,\theta^+_1\}, \quad \mathfrak{su}(1,1|2): \quad
\{\alpha^+_1,\beta^+_1,\theta^+_1,\eta^+_1\}, \quad \mathfrak{su}(2|3):
\quad \{\alpha^+_1,\alpha^+_2,\theta^+_1,\theta^+_2\}.  \nn \eea As we
can see there is a mixing between the sectors, the
$\mathfrak{su}(1|2)$ is contained within the larger
$\mathfrak{su}(2|3)$ sector and in $\mathfrak{su}(1,1|2)$, but the
latter is not a part of $\mathfrak{su}(2|3)$. When calculating the
energy shifts, things are straightforward for the first two sectors,
$\mathfrak{su}(1|2)$ and $\mathfrak{su}(1,1|2)$. The excited nodes are
$K_3$, $K_4$ and $K_5$ and for these excitation numbers
(\ref{constexpanded}) is significantly simplified since there are no
$u_{2,k}$ roots. Each $x_{3,k}$ and $x_{5,k}$ satisfy a $K_2-\nu$ degree
polynomial equation, where $\nu$ is the number of confluent mode numbers, 
which is the same for each value of $k$. However, this is not the case
for the $\mathfrak{su}(2|3)$ sector where we have nonvanishing $K_2$
excitations and a  resulting set of coupled polynomial equations for 
the $x_{2,k}$ and $x_{3,k}$ following from (\ref{constexpanded})

\subsection{The $\mathfrak{su}(1|2)$ sector}

As stated, this sector is spanned by the oscillators $\alpha^+_1$ and
$\theta^+_1$. The contributing parts from the string Hamiltonian are
$\mathcal{H}_{bb}$ and $\mathcal{H}_{bf}$. The explicit expression for the effective
$\mathfrak{su}(1|2)$ Hamiltonian can be found in
(\ref{eq:Hamiltonain_su(1|2)}).  Let us count the number of solutions for
the grading $\eta_1=\eta_2=1$.
Then the only excited nodes of the Dynkin
diagram in this sector are $K_4$ and $K_3$, so the polynomials in
(\ref{constexpanded}) give $K_4-\nu$ solutions\footnote{The number
  of confluent mode numbers must satisfy, $\nu \leq K_4-K_3+1$ since
  we cannot have fermionic excitations of the same flavor with
  confluent mode numbers.}. Two of these solutions are always $0$ and $
\infty$ while the other $K_4-2-\nu$ are non-trivial. Before we perform the actual
computation let us count the number of solutions. Say we have a total of $K_3$
 $\theta^+_1$ oscillators and $K_4-K_3$ $\alpha^+_1$
oscillators, then this state will yield $\frac{(K_4-\nu)\times
  (K_4-\nu-1)\times ... \times (K_4-\nu-K_3+1)}{K_3!}$ number of
solutions. So, for all possible combinations of a general $K_4$
impurity state the number of solutions are \bea
\label{numberofsolutions}
\sum_{K_3=0}^{K_4-\nu}\binom{K_4-\nu}{K_3}=2^{K_4-\nu}.
\eea
Since the worldsheet Hamiltonian is a $2^{K_4-\nu} \times 2^{K_4-\nu}$ matrix, the number 
of solutions matches.

\subsubsection{Two impurities}

For the two impurity sector the perturbative string Hamiltonian is a $4\times
4$ matrix, but we are only interested in a $2\times 2$ submatrix since
the other part falls into the rank one sectors $\mathfrak{su}(2)$ and
$\mathfrak{su}(1|1)$. The relevant matrix elements, with mode numbers
$\{q,-q\}$, are
\begin{displaymath}
\left(\begin{array}{c|c|c}
 & \alpha^+_{1,q}\theta^+_{1,-q} \ket{0} & \alpha^+_{1,-q}\theta^+_{1,q}\ket{0} \\
\hline
\bra{0} \alpha^-_{1,q} \theta^-_{1,-q} & \mathcal{H}_{bf} & \mathcal{H}_{bf}\\
\hline
\bra{0} \alpha^-_{1,-q} \theta^-_{1,q} & \mathcal{H}_{bf} & \mathcal{H}_{bf} 
\end{array}\right)
\end{displaymath}
The energy shifts are the non-zero values in
(\ref{eq:string_eigenvalues_su(1|2)_2imp}). Now, the interesting
question is of course if we can reproduce this result from the Bethe
equations. For the two impurity state $\alpha^+ \theta^+ \ket{0}$ it
is easiest to work with the gradings\footnote{All choices of gradings
  of course give the same result, however, the calculation will be
  more or less complicated depending on the choice.} $\eta_1=-1$ and
$\eta_2=1$ where we have $K_4=2$ and $K_3=1$. From
(\ref{constexpanded}) wee see that the only solutions for $x_{3,k}$
are 0 and $\infty$. Since we have two roots, and one $K_3$ excitation
we get two solutions for $p^1_k$. Solving (\ref{expandedbethe}) gives
$p^1_k=\pm p^0_k$. Plugging these into (\ref{energyshift}) gives \bea
\label{su12energy}
\delta P_- =\pm
\frac{\tilde{\lambda}}{P_+}\sum_{j=1}^2\frac{q_j^2}{\omega_{q_j}}=\pm
2\frac{\tilde{\lambda}}{P_+}\frac{q^2}{\omega_q}=:\kappa_2, \eea which equals
the non-zero values in (\ref{eq:string_eigenvalues_su(1|2)_2imp}).

\subsubsection{Three impurities, distinct mode numbers}

The full perturbative string Hamiltonian is a $8\times 8$ matrix but the
relevant $\mathfrak{su}(1|2)$ part splits up into two independent
submatrices coming from the Fermi-Fermi matrix elements 
$\bra{0}\alpha_1^- \alpha_1^-
\theta_1^- (\mathcal{H}_{bb}+\mathcal{H}_{bf}) \theta_1^+ \alpha_1^+
\alpha_1^+ \ket{0}$ and the Bose-Bose elements \\
$\bra{0}\alpha_1^- \theta_1^- \theta_1^-
(\mathcal{H}_{bf}) \theta_1^+ \theta_1^+ \alpha_1^+ \ket{0}$. Schematically
written we have,
\begin{equation}
\left(\begin{array}{c|c|c}
 & \alpha^+_1\alpha^+_1 \theta^+_1\ket{0} & \alpha^+_1 \theta^+_1 \theta^+_1\ket{0} \\
\hline
\bra{0} \theta^-_1 \alpha^-_1 \alpha^-_1 & (\mathcal{H}_{bb}+\mathcal{H}_{bf})^{3\times 3} & 0_{3\times 3}\\
\hline \bra{0} \theta^-_1\theta^-_1\alpha^-_1 & 0_{3\times 3} &
\mathcal{H}_{bf}^{3 \times 3}
\end{array}\right) \label{MAT1}
\end{equation}
The eigenvalues of the Bose-Bose submatrix, the bottom right, is given
in (\ref{eq:string_eigenvalues_su(1|2)_charge1}). To reproduce these
shifts from the Bethe equations we once again choose $\eta_1=-1$ and
$\eta_2=1$ so $K_4=3$ and $K_3=1$. Solving (\ref{constexpanded}) give,
as before, $x^0_{3,k}=\{0, \infty\}$ together with a novel third solution
\bea
\label{thirdsolutionsu12}
y=\frac{(2+\omega_{q_1}+\omega_{q_2})\,
  x^0_{4,3}+(2+\omega_{q_2}+\omega_{q_3})\,
  x^0_{4,1}+(2+\omega_{q_1}+\omega_{q_3})\,
  x^0_{4,2}}{3+\omega_{q_1}+\omega_{q_2}+\omega_{q_3}} \, .  \eea The
first two solutions, 0 and $\infty$, give as before $p^1_k=\pm
p^0_k$. For generic values of $K_4$, and with $K_3=1$, these two
solutions will always appear. Using the third solution in
(\ref{expandedbethe}) yields
\bea
\label{bossector}
p^1_k=\frac{1+\omega_{k}}{x^0_{4,k}-y}-p^0_k.
\eea
Plugging this into (\ref{energyshift}), together with some algebra, gives the three solutions
\bea
\label{2impsu12}
&& \delta P_-=\Big\{ \pm\frac{\tilde{\lambda}}{P_+}\sum_{j=1}^3\frac{q_j^2}{\omega_{q_j}}, ~ \frac{\tilde{\lambda}}{P_+ \omega_{q_1}\omega_{q_2}\omega_{q_3}}\sum_{j=1}^3q_j^2\omega_{q_j}\Big\}=:\Lambda_3\, ,
\eea
which agrees with the string result obtained in (\ref{eq:string_eigenvalues_su(1|2)_charge1}). 

Let us now focus on the Fermi-Fermi matrix elements, the upper left $3\times 3$ block of
(\ref{MAT1}). First,
(\ref{constexpanded}) give the same three solutions as before, namely
$\{0,\infty,y\}$ with the same $y$ as in
(\ref{thirdsolutionsu12}). Since $K_3=2$ we now, for each $p^1_k$, use
two of the solutions for $x^0_{3,k}$ \bea
p^1_k=(1+\omega_{p^0_k})\Big(\frac{1}{x^0_{4,k}-x^0_{3,1}}+\frac{1}{x^0_{4,k}-x^0_{3,2}}\Big)-2p^0_k.
\eea The three possible distributions of the roots, $\{0,\infty\},
\{0,y\}$ and $\{y,\infty\}$, give the three solutions
\bea
\label{3values}
\delta P_-=\Big\{ 0,~ - \frac{\tilde{\lambda}}{P_+}\frac{1}{16\pi^2}\sum_{j=1}^{K_4}\frac{p^0_k}{\omega_k}\Big((\frac{1+\omega_k}{x^0_{4,k}-y}-p^0_k)\pm p^0_k\Big) \Big\}=:\Omega_3
\eea
With a little bit of work one can show that these match the eigenvalues from the string Hamiltonian in (\ref{eq:string_eigenvalues_su(1|2)_charge2}).

\subsubsection{Three impurities, confluent mode numbers}

For three impurities, with mode numbers $\{q,q,-2q\}$, the only state
that does not fall into the already checked rank one sectors
\cite{ulcgauge} are $\alpha_{1}^+ \alpha_{1}^+ \theta_{1}^+ \ket{0}$
and $\alpha^+_1 \theta^+_1 \theta^+_1\ket{0}$. For the former, we get
from (\ref{expandedbethe}) (with grading $\eta_1=\eta_2=1$) \bea
{\tilde p}^2_q=-2p^0_q+\frac{2\omega_q+\omega_{2q}}{x^0_{4,q}-x^0_{4,2q}}-\frac{1+\omega_q}{x^0_{4,q}-x^0_3},
\quad
{\tilde p}^2_{2q}=-2p^0_{2q}+2\frac{2\omega_q+\omega_{2q}}{x^0_{4,2q}-x^0_{4,q}}-\frac{1+\omega_{2q}}{x^0_{4,2q}-x^0_3}. \nn
\eea The polynomials in (\ref{constexpanded}) give two solutions 
$\{0,\infty\}$ for $x^0_{3,k}$. Using these in (\ref{expandedenergycoinciding}),
together with some algebra, gives 
\bea
\label{coincidingsu12}
\delta P_-&=&
\frac{2q^2 \tilde{\lambda}}{P_+ \omega^2_q \omega_{2q}}\Big\{ 
    \frac{3\omega_{2q}+(2\omega_q+\omega_{2q})(4\omega_q(1+\omega_q)+\omega_{2q})}{3+2\omega_q+\omega_{2q}},
\nn\\ && \qquad\qquad \quad 
    -\frac{4\omega_q^2-(3-4\omega_q^2)\omega_{2q}-(1-2\omega_q)\omega_{2q}^2}{3+2\omega_q+\omega_{2q}}
  \Big\}.
\eea It is not immediately apparent that this equals the string
Hamiltonian result (\ref{su112conf2value}) but after some work one can
show that these two solutions are equal.

For the second state, $\alpha^+_1 \theta^+_1 \theta^+_1\ket{0}$, we
have $K_3=2$ and the two roots $\{0,\infty\}$ for $x^0_{3,k}$ can only be distributed
in one way. By doing analogously as above and using (\ref{expandedbethe})
in (\ref{expandedenergycoinciding}), we find \bea
\label{3impsu12onevalue}
\delta P_-= \frac{2q^2\tilde{\lambda}}{P_+}\frac{(\omega_q+\omega_{2q})}{\omega_q \omega_{2q}},
\eea
which reproduces the string Hamiltonian result of (\ref{su112conf1value}).

\subsection{The $\mathfrak{su}(1,1|2)$ sector}

Now we turn to the larger $\mathfrak{su}(1,1|2)$ sector. The procedure
is the same as above but now both sides of the Dynkin diagram gets
excited and a general state has the three middle nodes $K_3, K_4$ and $K_5$ excited. 
We are allowed to pick the same solution,
on the $K_3$ and $K_5$ node, but as before we must put distinct
solutions on the fermionic nodes. In this sector a new feature appears:
The states $\alpha^+_1 \beta^+_1$ and $\theta^+_1 \eta^+_1$ are allowed to 
mix. Also, in the case of confluent mode numbers, it
turns out that we have to make use of different gradings on some
states to generate all the solutions from the string Hamiltonian.

Let us first investigate if the number of solutions from the string
Hamiltonian and the Bethe equations match. A general
$\mathfrak{su}(1,1|2)$ state with $K_4$ excitations and $distinct$
mode numbers will yield a $2^{2K_4}\times 2^{2K_4}$ matrix and thus
$2^{2K_4}$ energy shifts. The total number of solutions from the Bethe
equations are just the square of (\ref{numberofsolutions}), with
$\nu=0$, which equals the number of eigenvalues from the perturbative
string Hamiltonian (\ref{eq:string_Hamiltonian_su(1|2)}).

\subsubsection{Two impurities}

The Hamiltonian is a $16\times 16$ matrix but it is only a $13\times
13$ part which lies outside the already calculated
$\mathfrak{su}(1|2)$ sector. There are seven different independent
submatrices where the largest is a $4 \times 4$ matrix and is
generated by the base kets $\alpha_1^+ \beta_1^+\ket{0}$ and $\theta_1^+
\eta_1^+ \ket{0}$. There are three $2 \times 2$ submatrices, $\alpha_1^+
\eta_1^+ \ket{0}, \beta_1^+ \theta_1^+ \ket{0}$ and $\beta_1^+ \eta_1^+
\ket{0}$. And three are one valued $\beta_1^+ \beta_1^+ \ket{0},\eta_1^+
\eta_1^+ \ket{0}$ and $\theta_1^+ \theta_1^+\ket{0}$, these will give the
same results as presented in \cite{ulcgauge} so these we will
ignore. The only part with mixing is the subpart generated by
$\alpha_1^+ \beta_1^+\ket{0}$ and $\theta_1^+ \eta_1^+ \ket{0}$. To calculate
the energy shifts we start by solving (\ref{constexpanded}) and, as
before, the two solutions are $\{0,\infty\}$. With $\eta_1=-1$ and
$\eta_2=1$, so $K_4=3$ and $K_5=K_3=1$, we have \bea
p^1_k=(1+\omega_{k})\Big(\frac{1}{x^0_{4,k}-x^0_{3,k}}-\frac{1}{x^0_{4,k}-x^0_{5,k}}\Big).
\eea Whenever we pick the same solution for $x^0_{3,k}$ and
$x^0_{5,k}$ we get zero and since we can do this in two ways we get
two zero solutions. The other two solutions are obtained by
setting $\{x^0_{3,k},x^0_{5,k}\}=\{0,\infty\}$ and $\{\infty,0\}$
which gives $p^1_k=\pm 2 p^0_k$. Using this in (\ref{energyshift})
gives \bea
\label{su1122imps}
\delta P_-=(0,0,\pm\frac{2
  \tilde{\lambda}}{P_+}\sum_{j=1}^2\frac{q_j^2}{\omega_{q_j}}), \eea
which is in agreement with the string Hamiltonian result in
(\ref{su1122impmixing}).

For the three parts $\alpha^+ \eta^+ \ket{0}, \beta^+ \theta^+
\ket{0}$ and $\beta^+ \eta^+ \ket{0}$, we see that solving for the
first state is analogous to the discussion after (\ref{su12energy}) but
with $\eta_1=1$ and $\eta_2=-1$. For the two other, the procedure will
again be identical if we choose the opposite gradings. That is, for
$\beta^+ \theta^+ \ket{0}$ we pick $\eta_1=1$ and $\eta_2=-1$, while
for $\beta^+ \eta^+\ket{0}$ we choose $\eta_1=-1$ and $\eta_2=1$ which
give the same set of solution for all three states \bea
\label{2impsu112}
\delta P_-=\pm\frac{2\tilde{\lambda}}{P_+}\frac{q^2}{\omega_q},
\eea
which is in agreement with (\ref{su1122imp}).

\subsubsection{Three impurities, distinct mode numbers}

The full perturbative string Hamiltonian will now be a $64\times 64$
matrix with non trivial $3\times 3$ and $9 \times 9$ subsectors. Since
the logic of solving the Bethe equation should be clear by now, 
we only present the obtained results in tabular form. Also, to make the
comparison with the string Hamiltonian more transparent, we now
also label the states by their charges $\{S_+,S_-,J_+,J_-\}$. 
The energy shifts for the $3 \times 3$ parts are given in
table \ref{33} and for the larger $9 \times 9$ subparts in
table \ref{99}. For the larger sectors we have a mixing between states
of different boson and fermion number.
\begin{table}[t]
\begin{center}
\begin{tabular}{|c|c|c|c|c|c|}
\hline
$\{\eta_1,\eta_2\}$ & $\{K_1+K_3,K_4,K_5+K_7\}$ & $\{S_+,S_-,J_+,J_-\}$ & $\delta P_-$  \\
\hline
$\{-,+\}$ & $\{2,3,0\}$  & $\{0,1,3,2\}_{\alpha^+_1\alpha^+_1\theta^+_1}$ & $\Omega_3$ \\
$\{+,-\}$ & $\{0,3,2\}$  & $\{1,0,2,3\}_{\alpha^+_1\alpha^+_1\eta^+_1}$ & $-\Omega_3$ \\
$\{-,+\}$ & $\{0,3,2\}$  & $\{2,3,1,0\}_{\beta^+_1\beta^+_1\theta^+_1}$ & $\Omega_3$ \\
$\{+,-\}$ & $\{2,3,0\}$  & $\{3,2,0,1\}_{\beta^+_1\beta^+_1\eta^+_1}$ & $-\Omega_3$ \\
\hline
$\{-,+\}$ & $\{1,3,0\}$  & $\{0,2,3,1\}_{\theta^+_1\theta^+_1\alpha^+_1}$ & $\Lambda_3$ \\
$\{-,+\}$ & $\{0,3,1\}$  & $\{1,3,2,0\}_{\theta^+_1\theta^+_1\beta^+_1}$ & $-\Lambda_3$ \\
$\{+,-\}$ & $\{0,3,1\}$  & $\{2,0,1,3\}_{\eta^+_1\eta^+_1\alpha^+_1}$ & $\Lambda_3$ \\
$\{+,-\}$ & $\{1,3,0\}$  & $\{3,1,0,2\}_{\eta^+_1\eta^+_1\beta^+_1}$ & $-\Lambda_3$ \\
\hline
\end{tabular}
\caption{The states reproducing the $3\times 3$ submatrices of the
  string Hamiltonian. $\Omega_3$ and $\Lambda_3$, where the subscript
  indicate the number of solutions as given in (\protect \ref{3values}) for
  $\Omega_3$ and (\protect \ref{2impsu12}) for $\Lambda_3$.}
\label{33}
\end{center}
\end{table}
\begin{table}[t]
\begin{center}
\begin{tabular}{|c|c|c|c|c|c|}
\hline
$\{\eta_1,\eta_2\}$ & $\{K_1+K_3,K_4,K_5+K_7\}$ & $\{S_+,S_-,J_+,J_-\}$ & $\delta P_-$  \\
\hline
$\{+,+\}$ & $\{1,3,1\}$  & $\{1,1,2,2\}_{(\alpha^+_1\alpha^+_1\beta^+_1),(\alpha^+_1\theta^+_1\eta^+_1)}$ & $\Omega_9$ \\
$\{-,-\}$ & $\{1,3,1\}$  & $\{2,2,1,1\}_{(\alpha^+_1\beta^+_1\beta^+_1),(\beta^+_1\theta^+_1\eta^+_1)}$ & $-\Omega_9$ \\
$\{-,+\}$ & $\{1,3,1\}$  & $\{1,2,2,1\}_{(\alpha^+_1\beta^+_1\theta^+_1),(\theta^+_1\theta^+_1\eta^+_1)}$ & $\Lambda_9$ \\
$\{+,-\}$ & $\{1,3,1\}$  & $\{2,1,1,2\}_{(\alpha^+_1\beta^+_1\eta^+_1,(\theta^+_1\eta^+_1\eta^+_1)}$ & $-\Lambda_9$ \\
\hline
\end{tabular}
\caption{The states reproducing the $9\times 9$ submatrices of the string Hamiltonian. $\Omega_9$ and $\Lambda_9$, where the subscript indicate the number of solutions, is given by (\protect \ref{3impbos}) and (\protect \ref{3impferm}).}
\label{99}
\end{center}
\end{table}

The functions $\Omega_9$ and $\Lambda_9$ in table \ref{99} depend on the mode numbers $\{q_1,q_2,q_3\}$ and are given by
\begin{align} \label{3impbos}
  \Omega_9 &= \frac{\tilde{\lambda}}{P_+ }\frac{1}{16\pi^2}\sum_{k=1}^{3}\frac{p^0_{q_k}}{\omega_{q_k}}
   	\Big(\sum_{j=1,j\neq k}^3\frac{2+\omega_{q_k}+\omega_{q_j}}{x^0_{4,q_k}-x^0_{4,q_j}}
			- \frac{1+\omega_{q_k}}{x^0_{4,q_k}-x^0_3}-\frac{1+\omega_{q_k}}{x^0_{4,q_k}-x^0_5})-p^0_{q_k}\Big)
\\
\label{3impferm}
\Lambda_9&=-\frac{\tilde{\lambda}}{P_+
}\frac{1}{16\pi^2}\sum_{k=1}^{3}\frac{p^0_{q_k}}{\omega_{q_k}}\Big(\frac{1+\omega_{q_k}}{x^0_{4,q_k}-x^0_3}-\frac{1+\omega_{q_k}}{x^0_{4,q_k}-x^0_5}\Big).
\end{align} 
To obtain the nine solutions for $\Omega_9$ and $\Lambda_9$  one has to insert one
of the three roots $\{0,\infty, y\}$ for each $x^0_3$ and
$x^0_5$. We have not managed to match these results with the perturbative string Hamiltonian
(\ref{eq:string_Hamiltonian_su(1|2)}) analytically, but tested the agreement extensively numerically.
The details of the numerical tests can be found in Appendix \ref{par:Numerics}.

\subsubsection{Three impurities, confluent mode numbers}

We will now look at three impurities with confluent mode numbers,
$\{q,q,-2q\}$. With two distinct mode numbers we see from
(\ref{constexpanded}) that we have the two standard solutions $\{0,\infty\}$
for $x^0_{3,k}$ and $x^0_{5,k}$.
\begin{table}[t]
\begin{center}
\begin{tabular}{|c|c|c|c|c|c|}
\hline
$\{\eta_1,\eta_2\}$ & $\{K_1+K_3,K_4,K_5+K_7\}$ & $\{S_+,S_-,J_+,J_-\}$ & $\delta P_-$  \\
\hline
$\{+,+\}$ & $\{1,3,0\}$  & $\{0,1,3,2\}_{\alpha^+_1\alpha^+_1\theta^+_1}$ & $\tilde{\Omega}_2$ \\
$\{+,+\}$ & $\{0,3,1\}$  & $\{1,0,2,3\}_{\alpha^+_1\alpha^+_1\eta^+_1}$ & $\tilde{\Omega}_2$ \\
$\{-,-\}$ & $\{0,3,1\}$  & $\{2,3,1,0\}_{\beta^+_1\beta^+_1\theta^+_1}$ & $-\tilde{\Omega}_2$ \\
$\{-,-\}$ & $\{1,3,0\}$  & $\{3,2,0,1\}_{\beta^+_1\beta^+_1\eta^+_1}$ & $-\tilde{\Omega}_2$ \\
\hline
$\{+,+\}$ & $\{2,3,0\}$  & $\{0,2,3,1\}_{\theta^+_1\theta^+_1\alpha^+_1}$ & $\tilde{\Lambda}_1$ \\
$\{-,-\}$ & $\{0,3,2\}$  & $\{1,3,2,0\}_{\theta^+_1\theta^+_1\beta^+_1}$ & $-\tilde{\Lambda}_1$ \\
$\{+,+\}$ & $\{0,3,2\}$  & $\{2,0,1,3\}_{\eta^+_1\eta^+_1\alpha^+_1}$ & $\tilde{\Lambda}_1$ \\
$\{-,-\}$ & $\{2,3,0\}$  & $\{3,1,0,2\}_{\eta^+_1\eta^+_1\beta^+_1}$ & $-\tilde{\Lambda}_1$ \\
\hline
\end{tabular}
\caption{The states reproducing the $2\times 2$ submatrices for
  $confluent$ mode numbers of the string
  Hamiltonian. $\tilde{\Omega}_2$ and $\tilde{\Lambda}_2$, where the
  subscript indicate the number of solutions, is given by
  (\protect \ref{coincidingsu12}) and (\protect \ref{3impsu12onevalue})}
\label{33conf}
\end{center}
\end{table}
The sectors exhibiting mixing, i.e.~the states that
span the $9 \times 9$ subparts of the previous section, now
exhibit a new feature. The gradings are no longer equivalent and
we will be forced to use both to generate all the desired
solutions. The simpler states, that do not exhibit this feature, are
presented in table\ref{33conf} and the states where different
gradings had to be used are presented in table \ref{99confluent}.
\begin{table}[t]
\begin{center}
\begin{tabular}{|c|c|c|c|c|c|}
\hline
$\{\eta_1,\eta_2\}$ & $\{K_1+K_3,K_4,K_5+K_7\}$ & $\{S_+,S_-,J_+,J_-\}$ & $\delta P_-$  \\
\hline
$\{+,+\}$ & $\{1,3,1\}$  & $\{1,1,2,2\}_{(\alpha^+_1\alpha^+_1\beta^+_1),(\alpha^+_1\theta^+_1\eta^+_1})$ & $\Gamma_4$ \\
$\{-,-\}$ & $\{2,3,2\}$  & $\{1,1,2,2\}_{(\alpha^+_1\alpha^+_1\beta^+_1),(\alpha^+_1\theta^+_1\eta^+_1})$ & $\tilde{\Gamma}_1$ \\
$\{-,-\}$ & $\{1,3,1\}$  & $\{2,2,1,1\}_{(\alpha^+_1\beta^+_1\beta^+_1),(\beta^+_1\theta^+_1\eta^+_1)}$ & $-\Gamma_4$ \\
$\{+,+\}$ & $\{2,3,2\}$  & $\{2,2,1,1\}_{(\alpha^+_1\beta^+_1\beta^+_1),(\beta^+_1\theta^+_1\eta^+_1)}$ & $-\tilde{\Gamma}_1$ \\
$\{+,+\}$ & $\{2,3,1\}$  & $\{1,2,2,1\}_{(\alpha^+_1\beta^+_1\theta^+_1),(\theta^+_1\theta^+_1\eta^+_1)}$ & $\tilde{\Omega}_2$ \\
$\{-,-\}$ & $\{1,3,2\}$  & $\{1,2,2,1\}_{(\alpha^+_1\beta^+_1\theta^+_1),(\theta^+_1\theta^+_1\eta^+_1)}$ & $-\tilde{\Omega}_2$ \\
$\{-,-\}$ & $\{2,3,1\}$  & $\{2,1,1,2\}_{(\alpha^+_1\beta^+_1\eta^+_1,(\theta^+_1\eta^+_1\eta^+_1)}$ & $-\tilde{\Omega}_2$ \\
$\{+,+\}$ & $\{1,3,2\}$  & $\{2,1,1,2\}_{(\alpha^+_1\beta^+_1\eta^+_1,(\theta^+_1\eta^+_1\eta^+_1)}$ & $\tilde{\Omega}_2$ \\
\hline
\end{tabular}
\caption{The states reproducing the larger submatrices, with {\sl confluent} mode numbers, of the string Hamiltonian. The functions $\Gamma_4$ and $\tilde{\Gamma}_1$ are given in (\protect \ref{gamma14}) and $\tilde{\Omega}_2$ is given in (\protect \ref{coincidingsu12}).}
\label{99confluent}
\end{center}
\end{table}
The energy shifts $\Gamma_4$ and $\tilde{\Gamma}_1$ appearing in table \ref{99confluent} are given by
\begin{align}
\label{gamma14}
\tilde{\Gamma}_1&=\frac{2 q^2 \tilde{\lambda}}{P_+ \omega_q^2 \omega_{2q}}\Big(\frac{1}{\omega_q}+\frac{1}{\omega_{2q}}\Big) \, , \nn\\ 
\Gamma_4&=-\frac{2 q^2 \tilde{\lambda}}{P_+ \omega_q^2 \omega_{2q}}\Big\{ (\frac{1}{\omega_q}+\frac{1}{\omega_{2q}}),(\frac{1}{\omega_q}+\frac{1}{\omega_{2q}}),
 \frac{3\omega_{2q}+(2\omega_q +\omega_{2q})(\omega_{2q}+\omega_q(7+6\omega_q+\omega_{2q}))}{3+2\omega_q+\omega_{2q}},
 \nn\\&
\frac{3\omega_{2q}-(2\omega_q +\omega_{2q})(\omega_q(5+2\omega_q+3\omega_{2q})-\omega_{2q})}{3+2\omega_q+\omega_{2q}}
\Big\}.
\end{align}
Again, for the comparison to the eigenvalues of the string Hamiltonian in this subsector we had
to resort to numerical verifications, see Appendix \ref{par:Numerics} for details.

\subsubsection{Higher impurities}
In going beyond three impurities numerical calculations on both sides, the Bethe equations and the string Hamiltonian,
have been performed for a number of four and five impurity states.
All numerical energy shifts match precisely, the tested configurations are listed 
in table \ref{tab:string_su(1,1|2)_higher_impurities}.
\begin{table}[t] \footnotesize
\begin{tabular}{|l |l l l |l |} \hline
%  \multicolumn{4}{l}{dimension $d=9$  } \\ \hline
$\{S_+,S_-,J_+,J_-\}$ & State pattern & & & Number of solutions \\ 
\hline
  $\{ 2,2,2,2 \}$ 	& $\theta^+_{1}\theta^+_{1}\eta^+_{1}\eta^+_{1}\ket{0}$,             & $\theta^+_{1}\eta^+_{1}\beta^+_{1}\alpha^+_{1}\ket{0}$,
  			& $\beta^+_{1}\beta^+_{1}\alpha^+_{1}\alpha^+_{1}\ket{0}$
  			& 36 energy shifts  \\
  $\{ 2,2,3,3 \}$ 	& $\theta^+_{1}\theta^+_{1}\eta^+_{1}\eta^+_{1}\alpha^+_{1}\ket{0}$, & $\theta^+_{1}\eta^+_{1}\beta^+_{1}\alpha^+_{1}\alpha^+_{1}\ket{0}$,
  			& $\beta^+_{1}\beta^+_{1}\alpha^+_{1}\alpha^+_{1}\alpha^+_{1}\ket{0}$
  			& 100 energy shifts  \\ \hline
\end{tabular}
\caption{Checked 4 and 5 impurity states of $\mathfrak{su}(1,1|2)$.}
\label{tab:string_su(1,1|2)_higher_impurities}
\end{table}

\subsection{The $\mathfrak{su}(2|3)$ sector}

Now things become more complex. The polynomials (\ref{constexpanded})
for a general state are highly non-linear, coupled and involve several
variables. For this reason we will not be as thorough in our testing for the higher
impurity cases as in the previous sections.
The oscillators in this sector are $\alpha^+_1,
\alpha^+_2, \theta^+_1$ and $\theta^+_2$ where there is a mixing
between $\alpha^+_1 \alpha^+_2\, |0\rangle$ and $\theta^+_1 \theta^+_2\, |0\rangle$. 
The string Hamiltonian is given in (\ref{eq:string_Hamiltonian_su(2|3)}).
\subsubsection{Two impurities}
The $\mathfrak{su}(2|3)$ two impurity sector of the perturbative
string Hamiltonian (\ref{eq:string_Hamiltonian_su(2|3)}) will be a $12
\times 12$ matrix. Let us begin with the largest subpart, the one with
mixing between $\alpha^+_1 \alpha^+_2 \ket{0}$ and $\theta^+_1
\theta^+_2 \ket{0}$.  The excitation numbers, with grading
$\eta_1=\eta_2=1$, for $\alpha^+_1 \alpha^+_2\ket{0}$ are
$K_1=K_2=K_3=1$ and $K_4=2$ while for $\theta^+_1 \theta^+_2 \ket{0}$
we have $K_2=1$ and $K_3=K_4=2$.  Here the dynamically transformed version of the Bethe equations
is advantageous, as it makes explicit that the relevant combination $K_1+K_3=2$ is the same for these
two states. This is how the Bethe equations take care of the mixing.
Solving for $u^0_2$ in
(\ref{constexpanded}), and using
$u^0_{3,k}=x^0_{3,k}+\frac{\tilde{\lambda}}{64\pi^2}\frac{1}{x^0_{3,k}}$,
gives \bea \nn
u^0_2=\frac{1}{2}(x^0_{3,1}+x^0_{3,2}+\frac{\tilde{\lambda}}{64\pi^2}(\frac{1}{x^0_{3,1}}+\frac{1}{x^0_{3,2}})).
\eea Plugging this into the second line of (\ref{constexpanded}) gives
\bea
\label{su23const}
&& \frac{1}{x^0_{3,1}-x^0_{3,2}+\frac{\tilde{\lambda}}{64 \pi^2}(\frac{1}{x^0_{3,1}}-\frac{1}{x^0_{3,2}})}+\sum_{j=1}^2\frac{1+\omega_{j}}{x^0_{4,j}-x^0_{3,1}}=0, \\ \nn
&& \frac{1}{x^0_{3,2}-x^0_{3,1}+\frac{\tilde{\lambda}}{64 \pi^2}(\frac{1}{x^0_{3,2}}-\frac{1}{x^0_{3,1}})}+\sum_{j=1}^2\frac{1+\omega_{j}}{x^0_{4,j}-x^0_{3,2}}=0.
\eea
We can add these two equations above and see that four
solutions are $(x^0_{3,1},x^0_{3,2})=(0,0),(0,\infty),(\infty,0)$ and
$(\infty, \infty)$. This may at first glance seem strange since the
seemingly equivalent state $\theta^+_1 \theta^+_2 \ket{0}$ only has the $K_2$ and
$K_3$ node excited, implying that we can not pick the same solution twice for
$x^0_{3,k}$ since $K_3$ is fermionic. However, the correct state to use is the $\alpha^+_1
\alpha^+_2\ket{0}$ state. Here two different fermionic nodes $K_1$ and $K_3$ are
excited and because of this we can use the same solutions on both
nodes simultaneously. 

Let us now turn to the calculation of the energy shifts for
the these four states. We use the solutions from
(\ref{su23const}) in (\ref{expandedbethe}) and plug this into
(\ref{energyshift}) which gives
\bea
\label{chi}
\delta P_-=\{0,0,\pm\frac{\tilde{\lambda}}{P_+}\frac{4
  q^2}{\omega_q}\}=:\chi_4, \eea which is in perfect agreement with
(\ref{su23mixingeigenvalue}). The energy shifts for the other states
follows immediately and we present the results in table \ref{22}. From
this table we see that all the energy shifts from
(\ref{eq:string_Hamiltonian_su(2|3)}), presented in
(\ref{su23eigenvalue}) and (\ref{su23mixingeigenvalue}), are
reproduced.
\begin{table}[t]
\begin{center}
\begin{tabular}{|c|c|c|c|c|c|}
\hline
$\{\eta_1,\eta_2\}$ & $\{K_1+K_3,K_2,K_4\}$ & $\{S_+,S_-,J_+,J_-\}$ & $\delta P_-$  \\
\hline
$\{+,+\}$ & $\{2,1,2\}$  & $\{0,0,2,0\}_{(\alpha^+_1\alpha^+_2),(\theta^+_1 \theta^+_2)}$ & $\chi_4$ \\
$\{-,+\}$ & $\{1,0,2\}$  & $\{0,1,2,1\}_{\alpha^+_1\theta^+_1}$ & $\kappa_2$ \\
$\{-,+\}$ & $\{1,0,2\}$  & $\{0,-1,2,-1\}_{\alpha^+_2\theta^+_2}$ & $\kappa_2$ \\
$\{+,+\}$ & $\{1,1,2\}$  & $\{0,-1,2,1\}_{\alpha^+_1\theta^+_2}$ & $\kappa_2$ \\
$\{+,+\}$ & $\{1,1,2\}$  & $\{0,1,2,-1\}_{\alpha^+_2\theta^+_1}$ & $\kappa_2$ \\
\hline
\end{tabular}
\caption{The two impurity states that fall into to the rank $\geq 1$ sectors for $\mathfrak{su}(2|3)$. Here $\chi_4$ is given by (\protect \ref{chi}) and $\kappa_2$ is given by (\protect \ref{su12energy}). For two of the states we have permutated the space-time indices.}
\label{22}
\end{center}
\end{table}
\subsubsection{Higher impurities}

Due to the non linearity of the polynomials relating the Bethe roots
we will only present results for excitations with $K_2=K_3=1$,
corresponding to states of the form $\alpha^+_1 \,\ldots\, \alpha^+_1
\theta^+_2 \ket{0}$ with space-time  charge vector $\{S_+,S_-,J_+,J_-\}=\{0,-1,K_4,K_4-1\}$. From the
first line in (\ref{constexpanded}) we see that \bea \nn
\frac{1}{u^0_2-(x^0_3+\frac{\tilde{\lambda}}{64
    \pi^2}\frac{1}{x^0_3})}=0, \eea and using this in the second line
implies that the equation for $x^0_3$ reduces to the familiar form
\bea \sum_{j=1}^{K_4}\frac{1+\omega_j}{x^0_{4,j}-x^0_3}=0.  \eea Thus,
the energy shift for this state is the same as for the $\alpha^+_1\,
... \, \alpha^+_1\theta^+_1 \ket{0}$ states. For $K_4=3$, the energy
shift is presented in (\ref{2impsu12}). For $K_4-1$ number of
$\,\alpha^+_1\,$ excitations and one $\,\theta^+_2 \,$ excitation, the
energy shift, with gradings $\{+,+\}$, is given by \bea
\label{higherimpenergy}
\Lambda_{K_4}=\frac{1}{16 \pi^2}\sum_{k=1}^{K_4}\frac{p^0_k}{\omega_k}\Big(\sum_{j=1 \atop j\neq k}^{K_4}\frac{2+\omega_j+\omega_k}{x^0_{4,k}-x^0_{4,j}}-\frac{1+\omega_k}{x^0_{4,k}-x^0_3}-p^0_k(K_4-1)\Big).
\eea
This prediction we have verified numerically for $K_4 \leq 6$ with the energy shifts obtained
by diagonalization of the string Hamiltonian (\ref{eq:string_Hamiltonian_su(2|3)}).
\begin{table}[t]
\begin{center}
\begin{tabular}{|c|c|c|c|c|c|}
\hline
$\{\eta_1,\eta_2\}$ & $\{K_1+K_3,K_2,K_4\}$ & $\{S_+,S_-,J_+,J_-\}$ & $\delta P_-$  \\
\hline
$\{+,+\}$ & $\{1,1,K_4\}$  & $\{0,-1,K_4,K_4-1\}_{(\alpha^+_1 \, ... \, \alpha^+_1\theta^+_2)}$ & $\Lambda_{K_4}$ \\
\hline
\end{tabular}
\caption{Higher impurity states from the $\mathfrak{su}(2|3)$ sector
  for states of the form $\alpha^+_1 \, ... \, \alpha^+_1 \theta^+_2
  \ket{0}$. The function $\Lambda_{K_4}$, where $K_4$ indicates the
  number of solutions, is given in (\protect \ref{higherimpenergy}).}
\end{center}
\end{table}

\section{Summary}

In this work we have explored the quantum integrability of the $AdS_5\times S^5$
superstring by confronting the conjectured set of Bethe equations with an explicit 
diagonalization of the light-cone gauged string Hamiltonian.

For this we have presented the Bethe equations 
for the most general excitation pattern of the uniform light-cone 
gauged $AdS_5\times S^5$ superstring in the near plane-wave limit. Moreover, it was
demonstrated how excited string states may be translated to 
distributions of spectral parameters in the 
Bethe equations as given in table 1. Using this we have explicity compared the predictions from the
light-cone Bethe equations with direct diagonalization of the string Hamiltonian in
perturbation theory at leading order in $1/P_+$. For
operators from the non dynamical sectors, we have verified the
spectrum for a large number of states giving us a strong confidence in
the validity of the light-cone Bethe equations for these classes of operators. 
For a generic $\mathfrak{su}(1,1|2)$ state, it is much easier to calculate the
energy shifts using the Bethe
equations. The characteristic polynomial from the perturbative string Hamiltonian
is of degree $2^{2K_4}$ wheras the polynomials needed to be solved in the Bethe equations
(\ref{constexpanded}) are of degree $K_4-2$. Still, one generically deals with polynomials of a
high degree, making it hard to explicitly find analytical results for states with large total excitation
number $K_4$.

When it comes to the dynamical sector $\mathfrak{su}(2|2)$, a direct
comparison is much more difficult due to the non linearity and coupled structure of
the Bethe equatons in (\ref{constexpanded}). Here analytical results were established
only for the two impurity case. Nevertheless, tests up to impurity number
six could be performed numerically.

In the light of this analysis it would be interesting to extend the perturbative study of the 
string Hamiltonian to next order in $1/P_+$. 
This is a very complicated problem due to normal ordering ambiguities. However,
this problem might be tackled by making use of the symmetry algebra as
discussed in \cite{ulcgauge} and \cite{offshell}. We hope to return to this issue
in the future.

\section*{Acknowledgements}

We wish to thank Sergey Frolov, Adam Rej, Matthias Staudacher and Tristan McLoughlin for valuable discussions. This work was supported by the Volkswagen-Foundation and the International Max-Planck
Research School for Geometric Analysis, Gravitation and String Theory.

% \pagebreak
%%%%%%%%%%%%%%%%%%%%%%%%%%%%%%%%%%%%%%%%%%%%%%%%%%%%%%%%%%%%%%%% Appendix %%%%%%%%%%%%%%%%%%%%%%%%%%%%%%%%%%%%%%%%%%%%%%%%%%%%%%%%%%%
\begin{appendix}
\section*{Appendix}
\section{Overview of the string results \label{par:Analytic_string_Results}}

To confront the proposed light-cone Bethe equations with the quantum string result
extensive computer algebra computations have been performed to diagonalize the 
worldsheed Hamiltonian perturbatively. For every considered subsector, i.e. 
$\mathfrak{su}(2)$, $\mathfrak{sl}(2)$,  $\mathfrak{su}(1|1)$, $\mathfrak{su}(1|2)$,
$\mathfrak{su}(1,1|2)$ and $\mathfrak{su}(2|3)$, we state the effective Hamiltonian and
present analytic results for its eigenvalues up to three impurities, whenever available. 
In some cases we had to retreat to a numerical comparison with the Bethe equations, 
details of these investigations are given in section \ref{par:Numerics}.

As one sees in table \ref{excpattern} the total number of impurities (or string excitations)
is given by $K_4$. We also allow for confluent mode numbers,
where the index $k=1,..,K'_4$ counts the excitations with distinct modes, each with a
multiplicity of $\nu_k$, using the notation of section 
\ref{section:Confluent_mode_numbers}. In uniform light-cone gauge the Hamiltonian
eigenvalue $-P_-$ is then given by
\begin{equation}
   P_- = -\sum^{K_4}_{k=1} \omega_k + \delta P_- = -\sum^{K'_4}_{k=1} \nu_k \hs{2pt}\omega_k + \delta P_-
\end{equation}
In order to classify the Hamiltonian eigenvalues we will make use of the $U(1)$ 
charges $\{S_+,S_-,J_+,J_- \}$ introduced in \cite{ulcgauge}. They
are light-cone combinations of the two spins $S_i$ of $AdS_5$ and two 
angular momenta $J_i$ on $S^5$,
viz.~$S_\pm=S_1\pm S_2$ and $J_\pm=J_1\pm J_2$. 
The charges of the string oscillators are spelled out in table 
\ref{tab:fermionic_and_bosonic_field_charges}.
\begin{table}[t] \footnotesize
  \begin{tabular}{|l|c|c|c|c|} \hline 
     							& $S_+$ & $S_-$ & $J_+$ & $J_-$ \\ \hline 
     $Y_1$, $P^y_1$, $\alpha^+_{1,m}$, $\alpha^-_{4,m}$ &  0   &  0   &  1  &  1 	\\
     $Y_2$, $P^y_2$, $\alpha^+_{2,m}$, $\alpha^-_{3,m}$ &  0   &  0   &  1  & -1	\\
     $Y_3$, $P^y_3$, $\alpha^+_{3,m}$, $\alpha^-_{2,m}$ &  0   &  0   & -1  &  1 	\\
     $Y_4$, $P^y_4$, $\alpha^+_{4,m}$, $\alpha^-_{1,m}$ &  0   &  0   & -1  & -1 	\\ \hline 
  \end{tabular} \hs{33pt}
  \begin{tabular}{|l|c|c|c|c|} \hline 
     							& $S_+$ & $S_-$ & $J_+$ & $J_-$ \\ \hline 
     $Z_1$, $P^z_1$, $\beta^+_{1,m}$, $\beta^-_{4,m}$ 	&  1   &  1   &  0  &  0 	\\
     $Z_2$, $P^z_2$, $\beta^+_{2,m}$, $\beta^-_{3,m}$ 	&  1   & -1   &  0  &  0 	\\
     $Z_3$, $P^z_3$, $\beta^+_{3,m}$, $\beta^-_{2,m}$ 	& -1   &  1   &  0  &  0 	\\
     $Z_4$, $P^z_4$, $\beta^+_{4,m}$, $\beta^-_{1,m}$ 	& -1   & -1   &  0  &  0 	\\ \hline 
  \end{tabular} \vs{3pt} \\
  \begin{tabular}{|l|c|c|c|c|} \hline 
     							& $S_+$ & $S_-$ & $J_+$ & $J_-$ \\ \hline 
     $\theta_1$, $\theta^\dagger_4$, $\theta^+_{1,m}$, $\theta^-_{4,m}$ &  0   &  1   &  1  &  0 	\\
     $\theta_2$, $\theta^\dagger_3$, $\theta^+_{2,m}$, $\theta^-_{3,m}$ &  0   & -1   &  1  &  0	\\
     $\theta_3$, $\theta^\dagger_2$, $\theta^+_{3,m}$, $\theta^-_{2,m}$ &  0   &  1   & -1  &  0 	\\
     $\theta_4$, $\theta^\dagger_1$, $\theta^+_{4,m}$, $\theta^-_{1,m}$ &  0   & -1   & -1  &  0 	\\ \hline 
  \end{tabular} \hs{46.3pt}
  \begin{tabular}{|l|c|c|c|c|} \hline 
     									& $S_+$ & $S_-$ & $J_+$ & $J_-$ \\ \hline 
     $\eta_1$, $\eta^\dagger_4$, $\eta^+_{1,m}$, $\eta^-_{4,m}$ 	&  1   &  0   &  0  &  1 	\\
     $\eta_2$, $\eta^\dagger_3$, $\eta^+_{2,m}$, $\eta^-_{3,m}$ 	&  1   &  0   &  0  & -1  	\\
     $\eta_3$, $\eta^\dagger_2$, $\eta^+_{3,m}$, $\eta^-_{2,m}$ 	& -1   &  0   &  0  &  1 	\\
     $\eta_4$, $\eta^\dagger_1$, $\eta^+_{4,m}$, $\eta^-_{1,m}$ 	& -1   &  0   &  0  & -1 	\\ \hline 
  \end{tabular} 
\caption{Charges of the annihilation and creation operators of the $AdS_5\times S^5$ string in
uniform light-cone gauge.}
\label{tab:fermionic_and_bosonic_field_charges}
\end{table}

\subsection{The $\mathfrak{su}(2)$ sector \label{par:String_su(2)_sector}}

This sector consists of states, which are composed only of
$\alpha^+_{1,n}$ creation operators. The Hamiltonian \eqref{hamiltonian} simplifies dramatically to the effective form
\begin{align} \label{eq:su(2)_Hamiltonian}
  \mathcal{H}_4^{(\mathfrak{su}(2))} = ~	
\widetilde{\lambda} \hs{-10pt} \sum_{{{\hs{5pt} m_1+m_2} \atop +m_3 + m_4 }= 0} 
 \frac{m_2 m_4}{\sqrt{\omega_{m_1} \omega_{m_2} \omega_{m_3} \omega_{m_4}}}
\, \alpha^+_{1,m_1} \alpha^+_{1,m_2} \alpha^-_{1,-m_3} \alpha^-_{1,-m_4} \hs{10pt}.
\end{align}
This sector is of rank one and the energy shifts $-\delta P_-$ for 
arbitrary modes $m_1,...,m_{K_4}$ can be evaluated to
\begin{align} \label{eq:string_eigenvalues_su(2)}
 \delta P_-^{(\mathfrak{su}(2))} 	=    \frac{\widetilde{\lambda}}{2P_+} 
\sum_{{i,j=1 \atop i\neq j}}^{K_4} \frac{(m_i+ m_j)^2}{\omega_{m_i} \omega_{m_j}}
- \frac{\widetilde{\lambda}}{P_+}\sum_{k=1}^{K_4'} \frac{m_k^2}{\omega_{m_k}^2} 
\nu_k \left( \nu_{k} -1\right)
\end{align}
By rewriting this $P_-$ shift in terms of the global energy $E$  and the BMN quantities 
$J$ and $\lambda' = \lambda/J^2$ using $P_\pm=J \pm E$, and then subsequently 
solving for $E$ one obtains the $\mathfrak{su}(2)$ global energy, which precisely 
agrees with the results in \cite{AFS} and 
\cite{N_I}
\begin{align} \label{eq:final_Energy} \notag
  E  &=   J +\sum_{k=1}^{K_4} \bar{\omega_k}
        - \frac{\lambda'}{4J}\sum_{k,j=1}^{K_4} \frac{m_k^2 \bar{\omega}_{j}^2  + m_j^2 \bar{\omega}_{k}^2}{\bar{\omega}_{k}\bar{\omega}_{j}}
        - \frac{\lambda'}{4J} \sum_{{i,j=1 \atop i\neq j}}^{K_4} \frac{(m_i+ m_j)^2}{\bar{\omega}_{i} \bar{\omega}_{j}}
        + \frac{\lambda'}{2J}\sum_{i=1}^{K'_4} \frac{m_i^2}{\bar{\omega}_{i}^2} \nu_k\left( \nu_{i} -1\right) \\
    & \tn{with } \hs{15pt} \bar{\omega}_k := \sqrt{1+\lambda'm_k^2} \, .
\end{align}

\subsection{The $\mathfrak{sl}(2)$ sector \label{par:String_sl(2)_sector}}

The $\mathfrak{sl}(2)$ states are composed of one flavor of
$\beta^+_{1,n}$ operators. Since the structure of the Hamiltonian is
identical for $\alpha^\pm_{1,n}$ and $\beta^\pm_{1,n}$ up to a minus
sign one immediately has
\begin{align}
  \mathcal{H}_4^{(\mathfrak{sl}(2))} & = ~- \widetilde{\lambda} \hs{-10pt} \sum_{{{\hs{5pt} m_1+m_2} \atop +m_3 + m_4 }= 0} 
  	\frac{m_2 m_4}{\sqrt{\omega_{m_1} \omega_{m_2} \omega_{m_3} \omega_{m_4}}} \, 
	\beta^+_{1,m_1} \beta^+_{1,m_2} \beta^-_{1,-m_3} \beta^-_{1,-m_4} \\
  \label{eq:string_eigenvalues_sl(2)}
  \delta P_-^{(\mathfrak{sl}(2))} & = - \delta P_-^{(\mathfrak{su}(2))} \,
\end{align}
and the global energy shift follows immediately.

\subsection{The $\mathfrak{su}(1\vert1)$ sector \label{par:String_su(1|1)_sector}}

States of the $\mathfrak{su}(1\vert1)$ sector are formed of
$\theta^+_{1,n}$ creation operators.  As noted in \cite{ulcgauge}
the restriction of the ${\cal O}(1/P_+)$ string Hamiltonian \eqref{hamiltonian} 
to the pure $\mathfrak{su}(1|1)$ sector vanishes
\begin{align}
  \mathcal{H}_{4}^{(\mathfrak{su}(1\vert 1))} \equiv 0 \hs{5pt}, \hs{20pt} \delta P_-^{(\mathfrak{su}(1\vert 1))} = 0 \hs{3pt} \, .
\end{align}

\subsection{The $\mathfrak{su}(1\vert 2)$ sector \label{par:String_su(1|2)_sector}}
We now turn to the first larger rank secor $\mathfrak{su}(1\vert2)$ being spanned by the creation
operators $\theta^+_{1,n}$ and $\alpha^+_{1,n}$ of one flavor. The effective Hamiltonian is
given by
\begin{align} \label{eq:Hamiltonain_su(1|2)}
 \mathcal{H}_{4}^{(\mathfrak{su}(1\vert 2))} = \mathcal{H}_4^{(\mathfrak{su}(2))}
	+\widetilde{\lambda} \hs{-10pt} \sum_{{{\hs{5pt} m_1+m_2} \atop +m_3 + m_4 }= 0}  \frac{X(m_1,m_2,m_3,m_4)}{\sqrt{\omega_{m_3} \omega_{m_4}}}  
        \hs{5pt} \theta_{1,m_1}^+ \theta_{1,-m_2}^- \alpha_{1,m_3}^+ \alpha_{1,-m_4}^-  \hs{10pt}.
\end{align}
where $X(m,n,k,l)$ is defined as
\begin{align} \label{eq:Defined_X_K_L_M_N}
       X(m,n,k,l) &:= \Big[ \Big( mn - \frac{(m-n)(k-l)}{4}\Big)(f_n f_m + g_n g_m) \nn\\
       &\quad - \frac{\kappa}{4\sqrt{\widetilde{\lambda}}}(k+l)(\omega_k +\omega_l) (f_n g_m + f_m g_n) \Big] \, ,
\end{align}
where $\kappa=\pm 1$.

\subsubsection{Two impurities}
For two impurity $\mathfrak{su}(1\vert2)$ states carrying the modes
$m_1=-m_2$ the Hamiltonian $\mathcal{H}_{4}$ forms a $4 \times 4$
matrix with eigenvalues $-\delta P_-$ where
\begin{align}
\label{eq:string_eigenvalues_su(1|2)_2imp}
  \delta P_- = \Big\{  \pm 2 \frac{\widetilde{\lambda}}{P_+}\frac{m_1^2}{\omega_1}, \hs{5pt} 0, \hs{5pt} 0 \Big\} \hs{5pt}.
\end{align}
\subsubsection{Three impurities with distinct modes}
Considering the three impurity case with distinct mode numbers
$m_1,m_2,m_3$ the Hamiltonian is represented by an $8 \times 8$ matrix
which decomposes into 4 non mixing submatrices, where two fall into the
rank one sectors $\mathfrak{su}(2)$ and $\mathfrak{su}(1\vert
1)$. The remaining pieces are two $3\times 3$ matrices.

Since string states only mix if they carry the same charges, we can classify the 
submatrices and their eigenvalues by the charge of the corresponding states. 
One finds:\\
$\{ S_+,S_-,J_+,J_- \} = \{ 0,2,3,1 \}_{\theta^+_{1} \theta^+_{1}
  \alpha^+_{1} \ket{0}}:$
\begin{align} \label{eq:string_eigenvalues_su(1|2)_charge1}
  \delta P_-= \Big\{ \pm \frac{ \widetilde{\lambda}}{P_+} \sum_{j=1}^3\frac{m_j^2}{\omega_j}, \hs{10pt} 
  		      \frac{\widetilde{\lambda}}{P_+ \omega_{1}\omega_{2}\omega_{3}}\sum_{j=1}^3 m_j^2 \hs{2pt} \omega_{j} \Big\}
\end{align}
$\{ S_+,S_-,J_+,J_- \} = \{ 0,1,3,2 \}_{\theta^+_{1} \alpha^+_{1} \alpha^+_{1} \ket{0}}:$
\begin{align} \label{eq:string_eigenvalues_su(1|2)_charge2}
 \delta P_- =&  \Big\{ 0, \hs{10pt} \frac{ \widetilde{\lambda} }{P_+}
	       \frac{m_1^2 \omega_{m_1} +m_2^2 \omega_{m_2} +m_3^2 \omega_{m_3} \pm \Xi_{m_1,m_2,m_3}}{ \omega_{m_1} \omega_{m_2} \omega_{m_3}} \Big\} \\ \notag
 \tn{with } \hs{10pt}  
 \Xi_{a,b,c} :=& \sqrt{ 4(\omega^2_{a}\chi^2_{b,c} + \omega^2_{b}\chi^2_{a,c} + \omega^2_{c}\chi^2_{a,b}) 
 			+ (\xi_{a;b,c} - \xi_{b;a,c}+\xi_{c;a,b})^2 -4 \xi_{a;b,c}\xi_{c;a,b} } \\ \notag
 \xi_{a;b,c} :=& -a(b\omega_b + c\omega_c  - a \omega_a ) \\ \notag
 \chi_{a,b} := & -ab\frac{\widetilde{\lambda}ab -(1+\omega_a)(1+\omega_b)}{\sqrt{(1+\omega_a)(1+\omega_b)}} \, .
\end{align}

\subsubsection{Three impurities with confluent modes}

In the case of confluent modes $\{ m_1, m_2, m_3 \} = \{m,m,-2m\}$ the submatrix with 
charges $\{ 0,2,3,1 \}$ collapses to a scalar whereas the submatrix of charge
 $\{ 0,1,3,2 \}$ reduces to $2\times 2$ matrix with energy shifts
\begin{flalign}
\label{su112conf1value}
  \{ S_+,S_-,J_+,J_- \} =  \{ 0,2,3,1 \}_{\theta^+_{1} \theta^+_{1} \alpha^+_{1} \ket{0}}: 
  	\hs{20pt}\delta P_- &= \frac{ \widetilde{\lambda} }{P_+} \frac{2m^2}{\omega_m}\big( \frac{1}{\omega_m} + \frac{1}{\omega_{2m}} \big) &
\end{flalign}
\begin{flalign}
\label{su112conf2value}
  & \{ S_+,S_-,J_+,J_- \} =  \{ 0,1,3,2 \}_{\theta^+_{1} \alpha^+_{1} \alpha^+_{1} \ket{0}}: &
\end{flalign} \vs{-20pt}
\begin{align*}
 \delta P_- = 2\frac{\tilde{\lambda}q^2}{P_+\omega_q^2\omega_{2q}}\Big(\omega_q+\omega_{2q}\pm\omega_q\sqrt{3+2\omega_{2q}^2+4\omega_q\omega_{2q}}\Big)
\end{align*}

\subsection{The $\mathfrak{su}(1,1\vert 2)$ sector \label{par:String_su(1,1|2)_sector}}

States of the $\mathfrak{su}(1,1\vert 2)$ sector are spaned by the set
$\{\theta^+_{1,n},\eta^+_{1,n},\beta^+_{1,n},\alpha^+_{1,n}\}$ of creation operators. In this sector the effective Hamiltonian takes the form 
\begin{align} \notag
   \mathcal{H}_4^{(\mathfrak{su}(1,1\vert 2))} = \hs{6pt} 
   	  & \widetilde{\lambda} \sum_{{k+l \atop +n+m}=0} \frac{kl}{\sqrt{\omega_m \omega_n \omega_k \omega_l}} 
	       (\alpha_{1,m}^+ \alpha_{1,-n}^- - \beta_{1,m}^+ \beta_{1,-n}^-)(\alpha_{1,k}^+ \alpha_{1,-l}^- + \beta_{1,k}^+ \beta_{1,-l}^-)\\ 
   \label{eq:string_Hamiltonian_su(1|2)}
        + & \widetilde{\lambda} \sum_{{k+l \atop +n+m}=0} 2\hs{2pt}\mathrm{i}\hs{1pt}\frac{f_m f_n - g_m g_n}{\sqrt{\omega_k \omega_l}}
     		(\theta_{1,m}^+\eta_{1,n}^+\beta_{1,-k}^-\alpha_{1,-l}^- + \theta_{1,-m}^-\eta_{1,-n}^-\beta_{1,k}^+\alpha_{1,l}^+) \\ \notag
	+ & \widetilde{\lambda} \sum_{{k+l \atop +n+m}=0} \frac{X(m,n,k,l)}{\sqrt{\omega_k \omega_l}}  
		(\theta_{1,m}^+\theta_{1,-n}^- + \eta_{1,m}^+\eta_{1,-n}^- )(\alpha_{1,k}^+\alpha_{1,-l}^- - \beta_{1,k}^+\beta_{1,-l}^- ) \hs{3pt},
\end{align}
where $X(m,n,k,l)$ is given in \eqref{eq:Defined_X_K_L_M_N}. 

\subsubsection{Two impurities}

The Hamiltonian matrix decomposes into several non mixing
submatrices. The $\mathfrak{su}(1,1\vert 2)$ sector contains all
previous discussed sectors, whose eigenvalues we do not state again. 
For the two impurity case with mode numbers $m_1 = -m_2$ one
obtains the new eigenvalues:
\begin{flalign}
\label{su1122impmixing}
  & \{ 1,1,1,1\}_{\theta^+_{1} \eta^+_{1} \ket{0},~ \beta^+_{1} \alpha^+_{1} \ket{0}}: 
    &\delta P_- &= \Big\{\pm 4 \frac{\widetilde{\lambda}}{P_+} \frac{m_1^2}{\omega_1} , \hs{5pt} 0, \hs{5pt} 0  \Big\} &  \\
\label{su1122imp}
  & \begin{array}{@{\vs{-3pt}}ll}
       \{1,2,1,0\}_{\theta^+_{1} \beta^+_{1} \ket{0}}, & \{0,1,2,1\}_{\theta^+_{1} \alpha^+_{1} \ket{0}} \\
       \{2,1,0,1\}_{\eta^+_{1} \beta^+_{1} \ket{0}},   & \{1,0,1,2\}_{\eta^+_{1} \alpha^+_{1} \ket{0}}
    \end{array}
     &\delta P_- &= \pm 2 \frac{\widetilde{\lambda}}{P_+} \frac{m_1^2}{\omega_1} & \\\nn
\end{flalign}

\subsubsection{Three impurities with confluent modes}

For higher impurities the situation becomes much more
involved. Already the three impurity $\mathfrak{su}(1,1\vert 2)$
Hamiltonian for non-confluent modes becomes a $64\times 64$ matrix
with submatrices of rank 9. We will classify the
$\mathfrak{su}(1,1\vert 2)$ submatrices with respect to their charges
and dimension $d$. Because $\mathfrak{su}(1,1\vert 2)$ contains
previously discussed sectors, we can deduce most of the eigenvalues by
using properties of the Hamiltonian
$\mathcal{H}_4^{(\mathfrak{su}(1,1\vert 2))}$. Our findings are collected in the
table \ref{tab:string_su(1,1|2)_3x3_matrices}.
\begin{table}[t] \footnotesize
\begin{tabular}{|l |l |l| l|}
  \multicolumn{4}{l}{dimension $d=1$  } \\ \hline
  $\{S_+,S_-,J_+,J_-\}$ 		& State pattern	& Property & $\delta P_-$ \\ \hline
  $\{ 0,0,3,3 \}$ 	& $\alpha^+_{1} \alpha^+_{1} \alpha^+_{1} \ket{0}$	&  $\mathfrak{su}(2)$ state 		& \eqref{eq:string_eigenvalues_su(2)} \\
  $\{ 3,3,0,0 \}$ 	& $\beta^+_{1} \beta^+_{1} \beta^+_{1} \ket{0}$		&  $\mathfrak{sl}(2)$ state 		& \eqref{eq:string_eigenvalues_sl(2)} \\
%  $\{ 0,3,3,0 \}$ 	& $\theta^+_{1} \theta^+_{1} \theta^+_{1} \ket{0}$		&  $\mathfrak{su}(1\vert 1)$ state	&  $\delta P_-^{\{ 0,3,3,0 \}} = 0$ \\
%  $\{ 3,0,0,3 \}$ 	& $\eta^+_{1}  \eta^+_{1}  \eta^+_{1} \ket{0}$		&  property of \eqref{hamiltonian} implies 
%  											&  $\delta P_-^{\{ 3,0,0,3 \}} = -\delta P_-^{\{ 0,3,3,0 \}} = 0$  \\ 
\hline \multicolumn{4}{l}{}\\ 
%\end{tabular}
%\caption{}
%\label{tab:string_su(1,1|2)_1x1_matrices}
%\end{table} 
%\begin{table}[h] \footnotesize
%\begin{tabular}{|l |l |l| l |}  
  \multicolumn{4}{l}{dimension $d=3$  } \\ \hline
$\{S_+,S_-,J_+,J_-\}$ 		& State pattern	& Property & $\delta P_-$ \\ \hline
  $\{ 0,2,3,1 \}$ 	& $\theta^+_{1} \theta^+_{1} \alpha^+_{1} \ket{0}$	&  $\mathfrak{su}(1\vert 2)$ state 
  											& $\delta P_-^{\{ 0,2,3,1 \}}$ see \eqref{eq:string_eigenvalues_su(1|2)_charge1} \\
  $\{ 2,0,1,3 \}$ 	& $\eta^+_{1} \eta^+_{1} \alpha^+_{1} \ket{0}$		&  property of \eqref{eq:string_Hamiltonian_su(1|2)} implies 
  											&  $\delta P_-^{\{ 2,1,0,3 \}} = +\delta P_-^{\{ 0,2,3,1 \}} $\\
  $\{ 1,3,2,0 \}$ 	& $\theta^+_{1} \theta^+_{1} \beta^+_{1} \ket{0}$	&  property of \eqref{eq:string_Hamiltonian_su(1|2)} implies 
  											&  $\delta P_-^{\{ 1,3,2,0 \}} = - \delta P_-^{\{ 0,2,3,1 \}}$ \\
  $\{ 3,1,0,2 \}$ 	& $\eta^+_{1} \eta^+_{1} \beta^+_{1} \ket{0}$		&  property of \eqref{eq:string_Hamiltonian_su(1|2)} implies 
  											&  $\delta P_-^{\{ 3,1,0,2 \}} = -\delta P_-^{\{ 0,2,3,1 \}} $\\
  $\{ 0,1,3,2 \}$ 	& $\theta^+_{1} \alpha^+_{1} \alpha^+_{1} \ket{0}$	&  $\mathfrak{su}(1\vert 2)$ state 
  											& $\delta P_-^{\{ 0,1,3,2 \}}$ see \eqref{eq:string_eigenvalues_su(1|2)_charge2} \\
  $\{ 1,0,2,3 \}$ 	& $\eta^+_{1} \alpha^+_{1} \alpha^+_{1} \ket{0}$	&  property of \eqref{eq:string_Hamiltonian_su(1|2)} implies 
  											&  $\delta P_-^{\{ 1,0,2,3 \}} = +\delta P_-^{\{ 0,1,3,2 \}} $\\
  $\{ 2,3,1,0 \}$ 	& $\theta^+_{1} \beta^+_{1} \beta^+_{1} \ket{0}$	&  property of \eqref{eq:string_Hamiltonian_su(1|2)} implies 
  											&  $\delta P_-^{\{ 2,3,1,0 \}} = -\delta P_-^{\{ 0,1,3,2 \}} $\\
  $\{ 3,2,0,1 \}$ 	& $\eta^+_{1} \beta^+_{1} \beta^+_{1} \ket{0}$		&  property of \eqref{eq:string_Hamiltonian_su(1|2)} implies 
  											&  $\delta P_-^{\{ 3,2,0,1 \}} = -\delta P_-^{\{ 0,1,3,2 \}} $ \\ \hline
\end{tabular}
\caption{Analytically accessible three impurity, distinct $\mathfrak{su}(1,1|2)$ energy shifts.}
\label{tab:string_su(1,1|2)_3x3_matrices}
\end{table}

The structure of the $9\times 9$ submatrices is a bit more involved.
Under the oscillator exchange $\theta_{1,m} \leftrightarrow \eta_{1,m}$ and 
$\alpha_{1,m} \leftrightarrow \beta_{1,m}$ 
%while leaving the normal order
%untouched\footnote{by convention we still consider for instance
%  $\theta_{1,m}^+\eta_{1,n}^+\beta_{1,-k}^-\alpha_{1,-l}^-$ as normal
%  ordered} 
the effective Hamiltonian $\mathcal{H}_4^{(\mathfrak{su}(1,1\vert 2))}$ changes its
sign. This exchange translates a state with charge $\{ 1,1,2,2 \}$ into one with
$\{ 2,2,1,1 \}$ or a $\{ 1,2,2,1 \}$ charged state into one with 
$\{2,1,1,2 \}$ and vice versa with mutual energy shifts of opposite signs.
See table \ref{tab:string_su(1,1|2)_9x9_matrices} for results.

\subsection{The $\mathfrak{su}(2\vert3)$ sector \label{par:String_su(2|3)_sector}}

Finally the $\mathfrak{su}(2\vert3)$ sector is spanned by the
operators $\theta^+_{1,n},\theta^+_{2,n}, \alpha^+_{1,n},
\alpha^+_{2,n}$. The effective form of $\mathcal{H}_4$ in this closed subsector reads 
\begin{align} \notag
  \mathcal{H}_4^{(\mathfrak{su}(2\vert 3))} = \hs{30pt} & \\ \notag
\label{eq:string_Hamiltonian_su(2|3)}
        \widetilde{\lambda} \sum_{{k+l \atop +n+m}=0} & \frac{kl}{\sqrt{\omega_m \omega_n \omega_k \omega_l}}
      		( \alpha_{1,m}^+ \alpha_{1,-n}^- + \alpha_{2,m}^+ \alpha_{2,-n}^-)(\alpha_{1,k}^+ \alpha_{1,-l}^- + \alpha_{2,k}^+ \alpha_{2,-l}^- ) \\
      + \widetilde{\lambda} \sum_{{k+l \atop +n+m}=0} & \frac{X(m,n,k,l)} {\sqrt{\omega_k \omega_l}}  
            	(\theta_{1,m}^+ \theta_{1,-n}^-+\theta_{2,m}^+ \theta_{2,-n}^-) ( \alpha_{1,k}^+ \alpha_{1,-l}^-  +\alpha_{2,k}^+ \alpha_{ 2,-l}^-)  \\  \notag
      - \frac{\widetilde{\lambda}}{2}\,\mathrm{i}\, \sum_{{k+l \atop +n+m}=0} &  \frac{1}{\sqrt{\omega_k \omega_l}}
	   	( \theta_{2,m}^+ \theta_{1,n}^+ \alpha_{2,-k}^- \alpha_{1,-l}^- + \theta_{2,-m}^- \theta_{1,-n}^- \alpha_{2,k}^+ \alpha_{1,l}^+ )  \\  \notag
        & {\times \left[ (m-n)(k-l)(f_n g_m - f_n g_m)  + \frac{\kappa}{\sqrt{\widetilde{\lambda}}}(k+l)(\omega_k -\omega_l) (f_n f_m - g_m g_n) \right] \atop } \\ \notag
+  \widetilde{\lambda} \sum_{{k+l \atop +n+m}=0} &
          \begin{pmatrix} 
	      \hs{8pt} (f_m g_n + f_n g_m)(f_k g_l + f_l g_k)(mn + kl) \\ + (f_n g_k + f_k g_n)(f_m g_l + f_l g_m)(nk + ml) \\
	      - (f_n f_l - g_n g_l)(f_m f_k + g_m g_k)(nl + mk)
	   \end{pmatrix}  \theta_{2,m}^+ \theta_{2,-n}^-  \theta_{1,k}^+ \theta_{1,-l}^- \, .
\end{align}
\begin{table}[t] \footnotesize
\begin{tabular}{|l |l l |l |}
  \multicolumn{4}{l}{dimension $d=9$  } \\ \hline
$\{S_+,S_-,J_+,J_-\}$ 		& State pattern	&  & $\delta P_-$ \\ \hline
  $\{ 1,1,2,2 \}$ 	& $\beta^+_{1} \alpha^+_{1} \alpha^+_{1} \ket{0}$, & $\theta^+_{1} \eta^+_{1} \alpha^+_{1} \ket{0}$	
  				&  rank 9 matrix, numerical eigenvalues see table \ref{tab:3_Impurity_first_order_corrections} \\
  $\{ 2,2,1,1 \}$ 	& $\beta^+_{1} \beta^+_{1} \alpha^+_{1} \ket{0}$, & $\theta^+_{1} \eta^+_{1} \beta^+_{1} \ket{0}$
  				&  $\delta P_-^{\{ 2,2,1,1 \}} = -\delta P_-^{\{ 1,1,2,2 \}} $\\
  $\{ 1,2,2,1 \}$ 	& $\theta^+_{1} \theta^+_{1} \eta^+_{1} \ket{0}$, & $\theta^+_{1} \beta^+_{1} \alpha^+_{1} \ket{0}$
  				&  rank 6 matrix, numerical eigenvalues see table \ref{tab:3_Impurity_first_order_corrections} \\
  $\{ 2,1,1,2 \}$ 	& $\theta^+_{1} \eta^+_{1} \eta^+_{1} \ket{0}$, & $\eta^+_{1} \beta^+_{1} \alpha^+_{1} \ket{0}$
  				&  $\delta P_-^{\{ 2,1,1,2 \}} = - \delta P_-^{\{ 1,2,2,1 \}} $ \\ \hline
\end{tabular}
\caption{Remaining three impurity, distinct $\mathfrak{su}(1,1|2)$ shifts, which were
compared numerically.}
\label{tab:string_su(1,1|2)_9x9_matrices}
\end{table}

\subsubsection{Two impurities}
For two impurities with mode numbers $m_2=-m_1$  we find the energy shifts
\begin{flalign}
\label{su23mixingeigenvalue}
  & \{ 0,0,2,0\}_{\theta^+_{2} \theta^+_{1} \ket{0}, ~\alpha^+_{2} \alpha^+_{1} \ket{0}}: 	 
  	& \hs{-30pt} \delta P_- &= \Big\{\pm 4\frac{\widetilde{\lambda}}{P_+} \frac{m_1^2}{\omega_1},~ 0, ~0 \Big\} & \\
\label{su23eigenvalue}
  & \begin{array}{ll}
       \{0,1,2,1\}_{\theta^+_{1} \alpha^+_{1} \ket{0}},   & \{0,1,2,-1\}_{\theta^+_{1} \alpha^+_{2} \ket{0}} \\ 
       \{0,-1,2,1\}_{\theta^+_{2} \alpha^+_{1} \ket{0}},  & \{0,-1,2,-1\}_{\theta^+_{2} \alpha^+_{2} \ket{0}}
    \end{array}
  &\delta P_- &= \pm 2 \frac{\widetilde{\lambda}}{P_+} \frac{m_1^2}{\omega_1} & \\\nn
\end{flalign}

\section{Numerical results\label{par:Numerics}}

Here we collect the numerical results, for this we dial explicit mode numbers and values for the
couping constant $\lambda'$. The considered cases constitute certain three impurity excitations
in the $\mathfrak{su}(1,1|2)$ subsector with distinct and confluent mode numbers, as well as
all three impurity excitations (distinct and confluent) for the $\mathfrak{su}(2|3)$ subsector. 
In the tables below we state explicit results for the values $\widetilde{\lambda}= 0.1$ 
and $P_+=100$ and mode numbers $(m_1,m_2,m_3)=\{(2,1,-3), (3,3,-6)\}$. All numerical energy
shifts were matched precisely with the result obtained from the Bethe
equations.
\begin{table}[t] \footnotesize
\begin{tabular}{|l| @{\hs{-17pt}} D{.}{.}{-1} @{\hs{-8pt}}D{.}{.}{-1} @{\hs{0pt}} D{.}{.}{-1} @{\hs{10pt}} D{.}{.}{-1} |}
\multicolumn{4}{l}{$\mathfrak{su}(2|3)$ sector\footnotemark  } \\ \hline
$\{S_+,S_-,J_+,J_-\}$ & \multicolumn{4}{c|}{eigenvalues $-\delta P_-$} \\ \hline 
\{0,0 ,3,$\pm 3$\}		& -0.0106324 & & & \\
\{0,$\pm 2$,3,$\pm 1$\}		& \pm 0.0108634 & -0.0106324 & & \\
\{0,$\pm 1$,3,$\pm 2$\}		& -0.0214958 & 0.000230962 & 0 & \\ 
\{0,$\pm 1$,3,0\}		& 0.0217267 & 3 \times -0.0214958 & 2\times 0.000230962 & 3 \times 0 \\ 
\{0,0,3,$\pm 1$\}		& -0.0323591 & 0.0110943   & 2\times \pm 0.0108634      & 3\times -0.0106324 \\ \hline 
\end{tabular} 

\begin{tabular}{|l| @{\hs{-17pt}}D{.}{.}{-1} @{\hs{-20pt}}D{.}{.}{-1} @{\hs{-7pt}}D{.}{.}{-1} @{\hs{7pt}}D{.}{.}{-1} @{\hs{-17pt}}D{.}{.}{-1} |}
\multicolumn{4}{l}{$\mathfrak{su}(1,1|2)$ sector  } \\ \hline
$\{S_+,S_-,J_+,J_-\}$ & \multicolumn{5}{c|}{eigenvalues $-\delta P_-$} \\ \hline 
\{1,1,2,2\}		& -0.0323591 & 0.0110943 & 2 \times \pm 0.0108634 & 2\times -0.0106324 &  0.0106324  \\
\{1,2,2,1\}, \{2,1,1,2\}& \pm 0.0217267 & \pm  0.0214958 & \pm 0.000230962 & 3 \times 0 & \\
\{2,2,1,1\}		& 0.0323591 & -0.0110943 & 2 \times  \pm 0.0108634 & 2\times 0.0106324 & -0.0106324 \\ \hline 
\end{tabular}
\caption{Numerical results for the first order correction in $1/P_+$ of the string energy spectrum for three impurity states with 
	distinct mode numbers $m_1 =2,  m_2 = 1, m_3 = -3$. 
	The number in front of some eigenvalues denotes their multiplicity if unequal to one. }
\label{tab:3_Impurity_first_order_corrections}
\end{table} \footnotetext{ The $\pm$ signs at some charges are just a shortform of writing several charge combinations all with the same eigenvalues. 
	They are not related to the signatures of the eigenvalues in any sense.}

\begin{table}[t] \footnotesize
\begin{tabular}{|l| D{.}{.}{-1} D{.}{.}{-1} D{.}{.}{-1} @{\hs{5pt}}D{.}{.}{-1} @{\hs{-5pt}}D{.}{.}{-1} |}
\multicolumn{4}{l}{$\mathfrak{su}(2|3)$ sector  } \\ \hline
$\{S_+,S_-,J_+,J_-\}$ & \multicolumn{5}{c|}{eigenvalues $-\delta P_-$} \\ \hline 
\{0,$\pm 1$,3,0\}	& 2\times -0.0454059 & 2\times 0.0142814 & & & \\
\{0,0,3,$\pm 1$\}	& -0.0752496 & 0.044125 & 3\times -0.0155623 & & \\
\{0,$\pm 2$,3,$\pm 1$\}, \{0,0,3,$\pm 3$\} & -0.0155623 &  &  & & \\
\{0,$\pm 1$,3,$\pm 2$\}	& -0.0454059 & 0.0142814  &  & & \\ \hline 
\end{tabular}

\begin{tabular}{|l| D{.}{.}{-1} D{.}{.}{-1} D{.}{.}{-1} @{\hs{5pt}}D{.}{.}{-1} @{\hs{-5pt}}D{.}{.}{-1} |}
\multicolumn{4}{l}{$\mathfrak{su}(1,1|2)$ sector  } \\ \hline
$\{S_+,S_-,J_+,J_-\}$  & \multicolumn{5}{c|}{eigenvalues $-\delta P_-$} \\ \hline 
\{1,1,2,2\}		& -0.0752496 & 0.044125 & 0.0155623 & 2\times -0.0155623 & \\
\{1,2,2,1\},\{2,1,1,2\}	& \pm 0.0454059 & \pm 0.0142814 & & & \\
\{2,2,1,1\}		& 0.0752496 & -0.044125 & 2 \times 0.0155623 & - 0.0155623 & \\ \hline 
\end{tabular} 
\caption{Numerical results for the first order correction in $1/P_+$ of the string energy spectrum for three impurity states 
	with confluent mode numbers $m_1 = m_2 = 3, m_3 = -6$. 
	The number in front of some eigenvalues denotes their multiplicity if unequal to one. }
\label{tab:3_Impurity_first_order_corrections_CONFLUENT}
\end{table}
\end{appendix}

\vfill
\newpage

%%%%%%%%%%%%%%%%%%%%%%%%%%%%%%%%%%%%%%%%%%%%%%%%%%%%%%%%%%%%%%%%%%%%%%%%%%%%%%%%%%%%%%%%%%%%%%%%%%%%%%%%%%%%%%%%%%%%%%%%%%%%%%%%%%%%%%%%%%%%%%%%%%%%%%%%


\begin{thebibliography}{99}
\bibitem{Metsaev:1998it}
  R.~R.~Metsaev and A.~A.~Tseytlin,
  ``Type IIB superstring action in $\AdS$ background,''
  Nucl.\ Phys.\  B {\bf 533} (1998) 109, hep-th/9805028.
  %%CITATION = NUPHA,B533,109;%%
\bibitem{ADSCFT}
%\cite{Maldacena:1997re}
  J.~M.~Maldacena,
  ``The large N limit of superconformal field theories and supergravity,''
  Adv.\ Theor.\ Math.\ Phys.\  {\bf 2}, 231 (1998)
  [Int.\ J.\ Theor.\ Phys.\  {\bf 38}, 1113 (1999)], hep-th/9711200. 
  %%CITATION = HEP-TH 9711200;%%
$\bullet$
  S.~S.~Gubser, I.~R.~Klebanov and A.~M.~Polyakov,
  ``Gauge theory correlators from non-critical string theory,''
  Phys.\ Lett.\ B {\bf 428} (1998) 105, hep-th/9802109.
  %%CITATION = HEP-TH 9802109;%%
$\bullet$
 E.~Witten,
  ``Anti-de Sitter space and holography,''
  Adv.\ Theor.\ Math.\ Phys.\  {\bf 2}, 253 (1998),
  hep-th/9802150.
  %%CITATION = HEP-TH 9802150;%%
%
\bibitem{MZ}
  J.~A.~Minahan and K.~Zarembo,
  ``The Bethe-ansatz for N = 4 super Yang-Mills,''
  JHEP {\bf 0303}, 013 (2003),
  hep-th/0212208.
  %%CITATION = HEP-TH 0212208;%%
%
\bibitem{revs}
 N.~Beisert,
``The dilatation operator of N = 4 super Yang-Mills theory and
integrability,''
Phys.\ Rept.\  {\bf 405} (2005) 1,
hep-th/0407277 
%%CITATION = HEP-TH 0407277;%%
$\bullet$
  K.~Zarembo,
  ``Semiclassical Bethe ansatz and AdS/CFT,''
  Comptes Rendus Physique {\bf 5} (2004) 1081
  [Fortsch.\ Phys.\  {\bf 53} (2005) 647],
  hep-th/0411191 $\bullet$
  %%CITATION = HEP-TH 0411191;%%
  J.~Plefka,
  ``Spinning strings and integrable spin chains in the AdS/CFT
  correspondence,'' Living Rev.~in Relativity 8, (2005), hep-th/0507136.
  %%CITATION = HEP-TH 0507136;%%
$\bullet$
 J.~A.~Minahan,
  ``A Brief Introduction To The Bethe Ansatz In N=4 Super-Yang-Mills,''
  J.\ Phys.\ A  {\bf 39} (2006) 12657.
  %%CITATION = JPAGB,A39,12657;%%

\bibitem{various}
 N.~Beisert, C.~Kristjansen and M.~Staudacher,
  ``The dilatation operator of N = 4 super Yang-Mills theory,''
  Nucl.\ Phys.\ B {\bf 664} (2003) 131,
  hep-th/0303060.
  %%CITATION = HEP-TH 0303060;%%
$\bullet$
 N.~Beisert and M.~Staudacher,
  ``The N = 4 SYM integrable super spin chain,''
  Nucl.\ Phys.\  B {\bf 670} (2003) 439,
  hep-th/0307042`.
  %%CITATION = NUPHA,B670,439;%%

\bibitem{Beisert:2005di}
  N.~Beisert, V.~A.~Kazakov, K.~Sakai and K.~Zarembo,
  ``Complete spectrum of long operators in N = 4 SYM at one loop,''
  JHEP {\bf 0507} (2005) 030, hep-th/0503200.
  %%CITATION = JHEPA,0507,030;%%

\bibitem{longrange}
  N.~Beisert and M.~Staudacher,
  ``Long-range PSU(2,2$|$4) Bethe ans\"atze for gauge theory and strings,''
  Nucl.\ Phys.\ B {\bf 727} (2005) 1,
  hep-th/0504190.
  %%CITATION = HEP-TH 0504190;%%

\bibitem{Staudacher:2004tk}
  M.~Staudacher,
  ``The factorized S-matrix of CFT/AdS,''
  JHEP {\bf 0505} (2005) 054, hep-th/0412188.
  %%CITATION = JHEPA,0505,054;%%

\bibitem{B0511}
  N.~Beisert,
  ``The su(2$|$2) dynamic S-matrix,''
  hep-th/0511082.
  %%CITATION = HEP-TH/0511082;%%

\bibitem{Janik}
  R.~A.~Janik,
  ``The $\AdS$ superstring worldsheet S-matrix and crossing symmetry,''
  Phys.\ Rev.\ D {\bf 73} (2006) 086006, hep-th/0603038.
  %%CITATION = HEP-TH 0603038;%%

\bibitem{Hopf}
C.~Gomez and R.~Hernandez,
  ``The magnon kinematics of the AdS/CFT correspondence,''
  JHEP {\bf 0611} (2006) 021,
  hep-th/0608029.
  %%CITATION = JHEPA,0611,021;%%
$\bullet$
 J.~Plefka, F.~Spill and A.~Torrielli,
  ``On the Hopf algebra structure of the AdS/CFT S-matrix,''
  Phys.\ Rev.\  D {\bf 74} (2006) 066008,
  hep-th/0608038.
  %%CITATION = PHRVA,D74,066008;%%

\bibitem{BHLBES}
N.~Beisert, R.~Hernandez and E.~Lopez,
  ``A crossing-symmetric phase for $\AdS$ strings,''
  JHEP {\bf 0611} (2006) 070,
  hep-th/0609044.
  %%CITATION = JHEPA,0611,070;%%
$\bullet$
N.~Beisert, B.~Eden and M.~Staudacher,
  ``Transcendentality and crossing,''
  J.\ Stat.\ Mech.\  {\bf 0701} (2007) P021, hep-th/0610251.
  %%CITATION = JSTAT,0701,P021;%%

\bibitem{4loops}
Z.~Bern, M.~Czakon, L.~J.~Dixon, D.~A.~Kosower and V.~A.~Smirnov,
  ``The four-loop planar amplitude and cusp anomalous dimension in maximally
  supersymmetric Yang-Mills theory,''
  hep-th/0610248.
  %%CITATION = HEP-TH/0610248;%%
$\bullet$
F.~Cachazo, M.~Spradlin and A.~Volovich,
  ``Four-loop cusp anomalous dimension from obstructions,''
  hep-th/0612309.
  %%CITATION = HEP-TH/0612309;%%

\bibitem{Bena:2003wd}
I.~Bena, J.~Polchinski and R.~Roiban, ``Hidden symmetries of the
$\AdS$ superstring,'' Phys.\ Rev.\ D {\bf 69} (2004) 046002,
hep-th/0305116.
%%CITATION = HEP-TH 0305116;%%

%\cite{Hofman:2006xt}
\bibitem{Hofman:2006xt}
  D.~M.~Hofman and J.~M.~Maldacena,
  ``Giant magnons,''
  J.\ Phys.\ A  {\bf 39} (2006) 13095,
  hep-th/0604135.
  %%CITATION = JPAGB,A39,13095;%%

\bibitem{BMN}
  D.~Berenstein, J.~M.~Maldacena and H.~Nastase,
  ``Strings in flat space and pp waves from N = 4 super Yang Mills,''
  JHEP {\bf 0204}, 013 (2002),
  hep-th/0202021.
  %%CITATION = HEP-TH 0202021;%%
\bibitem{Parnachev}
A.~Parnachev and A.~V.~Ryzhov,
  ``Strings in the near plane wave background and AdS/CFT,''
  JHEP {\bf 0210} (2002) 066, hep-th/0208010.
  %%CITATION = JHEPA,0210,066;%%
\bibitem{Callanetal}
C.~G.~Callan, H.~K.~Lee, T.~McLoughlin, J.~H.~Schwarz, I.~Swanson
and X.~Wu, ``Quantizing string theory in $\AdS$: Beyond the
pp-wave,''
 Nucl.\ Phys.\ B {\bf 673} (2003) 3,
hep-th/0307032. \,
%%CITATION = HEP-TH 0307032;%%
$\bullet$ 
C.~G.~Callan, T.~McLoughlin and I.~Swanson, ``Holography
beyond the Penrose limit,'' Nucl.\ Phys.\ B {\bf 694} (2004) 115,
hep-th/0404007. \,
%%CITATION = HEP-TH 0404007;%%
$\bullet$ 
C.~G.~.~Callan, T.~McLoughlin and I.~J.~Swanson,
  ``Higher impurity AdS/CFT correspondence in the near-BMN limit,''
  Nucl.\ Phys.\  B {\bf 700} (2004) 271, hep-th/0405153].
  %%CITATION = NUPHA,B700,271;%%

\bibitem{N_I}
T.~McLoughlin and I.~J.~Swanson,
  ``N-impurity superstring spectra near the pp-wave limit,''
  Nucl.\ Phys.\  B {\bf 702} (2004) 86, hep-th/0407240.
  %%CITATION = NUPHA,B702,86;%%

\bibitem{Martins:2007hb}
  M.~J.~Martins and C.~S.~Melo,
  ``The spectrum of particles interacting through centrally extended su(2$|$2)
  S-matrices,'' hep-th/0703086.
  %%CITATION = HEP-TH/0703086;%%

\bibitem{AFS}
  G.~Arutyunov, S.~Frolov and M.~Staudacher,
   ``Bethe ansatz for quantum strings,''
  JHEP {\bf 0410}, 016 (2004), hep-th/0406256.
  %%CITATION = HEP-TH 0406256;%%

\bibitem{ulc1}
G.~Arutyunov and S.~Frolov,
  ``Integrable Hamiltonian for classical strings on $\AdS$,''
  JHEP {\bf 0502} (2005) 059, hep-th/0411089.
  %%CITATION = JHEPA,0502,059;%%

\bibitem{ulc2} 
 G.~Arutyunov and S.~Frolov,
  ``Uniform light-cone gauge for strings in $\AdS$: Solving su(1$|$1)
  sector,''
  JHEP {\bf 0601} (2006) 055, 
  hep-th/0510208.
  %%CITATION = JHEPA,0601,055;%%

\bibitem{ulcgauge} 
S.~Frolov, J.~Plefka and M.~Zamaklar,
  ``The $\AdS$ superstring in light-cone gauge and its Bethe
  equations,''
  J.\ Phys.\ A  {\bf 39} (2006) 1303, hep-th/0603008.
  %%CITATION = JPAGB,A39,13037;%%

\bibitem{offshell} 
G.~Arutyunov, S.~Frolov, J.~Plefka and M.~Zamaklar,
  ``The off-shell symmetry algebra of the light-cone $\AdS$
  superstring,''
  hep-th/0609157.
  %%CITATION = HEP-TH/0609157;%%

\bibitem{ZF}
 G.~Arutyunov, S.~Frolov and M.~Zamaklar,
  ``The Zamolodchikov-Faddeev algebra for $AdS_5\times S^5$ superstring,''
  hep-th/0612229.
  %%CITATION = HEP-TH/0612229;%%

\bibitem{McLKRZ}
  T.~Klose, T.~McLoughlin, R.~Roiban and K.~Zarembo,
  ``Worldsheet scattering in $\AdS$,''
  hep-th/0611169.
  %%CITATION = HEP-TH/0611169;%%

%\cite{Klose:2006dd}
\bibitem{KloseZarembo}
  T.~Klose and K.~Zarembo,
  ``Bethe ansatz in stringy sigma models,''
  J.\ Stat.\ Mech.\  {\bf 0605} (2006) P006,
  hep-th/0603039.
  %%CITATION = JSTAT,0605,P006;%%
$\bullet$
  T.~Klose and K.~Zarembo,
  ``Reduced sigma-model on $\AdS$: One-loop scattering amplitudes,''
  JHEP {\bf 0702} (2007) 071,
  hep-th/0701240.
  %%CITATION = JHEPA,0702,071;%%

%\cite{Arutyunov:2006iu}
\bibitem{Arutyunov:2006iu}
  G.~Arutyunov and S.~Frolov,
  ``On $\AdS$ string S-matrix,''
  Phys.\ Lett.\  B {\bf 639} (2006) 378,
  hep-th/0604043.
  %%CITATION = PHLTA,B639,378;%%


\bibitem{nested} C.-N. Yang, "Some exact results for the many body
  problems in one dimension with repulsive delta function
  interaction," Phys. Rev Lett. 19, 1312 (1967).

\bibitem{stieltjes} T. J. Stieltjes. "Sur Quelques Theoremes
  d'Algebre, "Ouvres Completes. vol. 1, p 440, Noordhoff, Groningen,
  The Netherlands (1914)
$\bullet$
B. S. Shastry. A. Dhar, "Solution of a
  generalized Stieltjes Problem" J.~Phys~A~{\bf 34} (2001) 6197-6208, cond-mat/0101464.



\end{thebibliography}
\end{document}